\documentclass[preprint,12pt]{elsarticle}
\usepackage{amssymb}
\usepackage{lineno,hyperref}
\usepackage{hyperref}
\usepackage{fancyhdr}
\usepackage{amsmath}
\usepackage{mathtools}
\usepackage{amsfonts}
\usepackage{amssymb}
\usepackage{textcomp}
\usepackage{booktabs}                      
\usepackage{multicol}
\usepackage{algorithm}
\usepackage{algpseudocode}
\usepackage{subcaption,booktabs}
\usepackage{multirow}
\usepackage{nomencl}
\usepackage{caption}
\usepackage{float}
\usepackage{rotating}
\usepackage{tabularx}                      
\usepackage{natbib}
\usepackage{adjustbox}
\usepackage[graphicx]{realboxes}
\usepackage[toc,page]{appendix}
\usepackage{cancel}
\usepackage{graphicx}
\usepackage{color}
\usepackage{xcolor}
\usepackage{cleveref}
\crefformat{section}{\S#2#1#3} 
\crefformat{subsection}{\S#2#1#3}
\crefformat{subsubsection}{\S#2#1#3}
\newcounter{bla}

\def\code#1{\texttt{#1}}

\journal{Computer Physics Communications}

\begin{document}

\begin{frontmatter}



\title{FluTAS: A GPU-accelerated finite difference code for multiphase flows}

%
%
\author[a]{Marco Crialesi-Esposito\corref{a1}\corref{a2}}\ead{marcoce@kth.se}
\author[a]{Nicol\`o Scapin\corref{a1}}\ead{nicolos@mech.kth.se}
\author[b]{Andreas D. Demou}\ead{demou@mech.kth.se}
\author[c]{Marco Edoardo Rosti}\ead{marco.rosti@oist.jp}
\author[d]{Pedro Costa}\ead{pcosta@hi.is}
\author[e]{Filippo Spiga}\ead{fspiga@nvidia.com}
\author[a,f]{Luca Brandt}\ead{luca@mech.kth.se}

\cortext[a1]{M.C-E. and N.S. contributed equally to this work.}
\cortext[a2]{Corresponding author.}
\address[a]{Department of Engineering Mechanics, Royal Institute of Technology (KTH), Stockholm, Sweden,}
\address[b]{The Cyprus Institute, Nicosia, Cyprus,}
\address[c]{Complex Fluids and Flows Unit, Okinawa Institute of Science and Technology Graduate University (OIST), 1919-1 Tancha, Onna-son, Okinawa 904-0495, Japan,}
\address[d]{Faculty of Industrial Engineering, Mechanical Engineering and Computer Science, University of Iceland, Hjardarhagi 2-6, 107 Reykjavík, Iceland}
\address[e]{NVIDIA Ltd, Cambridge (UK),}
\address[f]{Department of Energy and Process Engineering, Norwegian University of Science and Technology (NTNU), Trondheim, Norway.}

\begin{abstract}
We present the Fluid Transport Accelerated Solver, FluTAS, a scalable GPU code for multiphase flows with thermal effects. The code solves the incompressible Navier-Stokes equation for two-fluid systems, with a direct FFT-based Poisson solver for the pressure equation. The interface between the two fluids is represented with the Volume of Fluid (VoF) method, which is mass conserving and well suited for complex flows thanks to its capacity of handling topological changes. The energy equation is explicitly solved and coupled with the momentum equation through the Boussinesq approximation. The code is conceived in a modular fashion so that different numerical methods can be used independently, the existing routines can be modified, and new ones can be included in a straightforward and sustainable manner. FluTAS is written in modern Fortran and parallelized using hybrid MPI/OpenMP in the CPU-only version and accelerated with OpenACC directives in the GPU implementation. We present different benchmarks to validate the code, and two large-scale simulations of fundamental interest in turbulent multiphase flows: isothermal emulsions in HIT and two-layer Rayleigh-B\'enard convection. FluTAS is distributed through a MIT license and arises from a collaborative effort of several scientists, aiming to become a flexible tool to study complex multiphase flows.
\end{abstract}


\end{frontmatter}


\noindent
{\bf Program summary} \\

\begin{small}
\noindent
{\em Program Title}: \textit{Fluid Transport Accelerated Solver}, FluTAS. \\
{\em Developer's repository link:} \url{https://github.com/Multiphysics-Flow-Solvers/FluTAS.git}. \\
{\em Licensing provisions:} MIT License. \\
{\em Programming language:} Fortran 90, parallelized using MPI and slab/pencil decomposition, GPU accelerated using OpenACC directives. \\
{\em External libraries/routines:} FFTW, cuFFT. \\
{\em Nature of problem:} FluTAS is a GPU-accelerated numerical code tailored to perform interface resolved simulations of incompressible multiphase flows, optionally with heat transfer. The code combines a standard pressure correction algorithm with an algebraic volume of fluid method, MTHINC~\cite{ii2012interface}. \\
{\em Solution method:} the code employs a second-order-finite difference discretization and solves the two-fluid Navier-Stokes equation using a projection method. It can be run both on CPU-architectures and GPU-architectures.

\end{small}

\section{Introduction}\label{sec:intro} 
Multiphase flows are ubiquitous in many contexts, ranging from environmental flows to industrial applications. The interaction between phases has a prominent role in the formation and evolution of clouds~\cite{grabowski2013growth}, in sediment transport~\cite{seminara2010fluvial,brandt2021particle}, oceanic sprays and bubbles generation~\cite{veron2015ocean} and more in general it represents one of the \textit{grand challenges of environmental fluid mechanics}~\cite{dauxois2021confronting}. These flows are also crucial in several industrial applications, such as pharmaceutical, transportation, food processing and power generation~\cite{crowe2005multiphase}. From a theoretical point of view, the main difficulty when analyzing multiphase flows relies on their multiscale nature, since the length-scale of the interface is of the order of the mean-free path while in most applications the typical length-scale is several orders of magnitude larger ($\sim 10^5-10^6$). This huge separation of scales is magnified when dealing with turbulent multiphase flows, which further broadens the spectrum of length-scales, thus making unfeasible any attempt to bridge all of them in a single and unique framework. For this reason, all the tools developed so far, both of experimental and numerical nature, have focused on only a portion of the scale spectrum while the remaining part is modelled or neglected. \par
As regards multiphase turbulence, where most of our interests and applications are, both experimental investigations and numerical simulations have been extensively used in the last thirty years and have led to important contributions in a variety of problems and configurations: to name a few, particle laden flows and sediment transport discussed in~\cite{voth2017anisotropic,brandt2021particle}, bubbly and droplet flows reviewed in~\cite{risso2018agitation,elghobashi2019direct,mathai2020bubbly}, oceanic sprays and bubbles as highlighted in~\cite{veron2015ocean}. Nevertheless, as already discussed in~\cite{elghobashi2019direct} and despite the recent progress in the instantaneous measurement of bubble/droplet shape~\citep{masuk2019robust,salibindla2020lift,masuk2021simultaneous}, there is still a lack of experimental data for the measurement of the instantaneous velocity fields of both carried and dispersed phase as well as for the turbulent kinetic energy and dissipation near the interface locations. 
These limitations disappear when dealing with numerical simulations and, therefore, in the last decades interface resolved simulations of multiphase flows have become a central investigation tool. Despite the advantages, numerical simulations are still limited to simple configurations and moderate scale separation, and pose the challenge of the choice of the proper method to fully resolve the two-phase interface. As discussed in~\cite{mirjalili2017interface}, there is now consensus that numerical methods suitable to perform interface-resolved simulations of multiphase flow should have the following properties: i) be able to enforce mass, momentum and kinetic energy conservation at discrete level, ii) allow mismatches in the material properties, whose magnitude depends on the application, and iii) handle complex and possibly arbitrary topological changes. Among the four groups of numerical methods for multiphase flows, Front-Tracking (FT)~\cite{unverdi1992front}, Volume-of-Fluid (VoF)~\cite{scardovelli1999direct}, Phase Field (PFM)~\cite{anderson1998diffuse}, Level-set (LS)~\cite{sethian2003level}, it exists at least a variant of each which possesses the aforementioned numerical properties, giving some freedom to researchers and scientists on the choice of their preferred numerical tool (see~\cite{scardovelli1999direct,prosperetti2009computational,soligo2021turbulent} for a review). \par
Nevertheless, it is becoming ever more clear that another desirable property of any numerical method is its straightforward adaptation to be able to run massively parallel simulations, especially on accelerated architectures. With the increase in the computing power driven by Graphics Processing Units (GPUs)~\cite{khan2021analysis}, several HPC centers are now shifting towards GPU-only and GPU-accelerated architectures. This trend is making the GPU-parallelization of numerical codes for fluid mechanics a mandatory requirement rather than a simple advantage. This effort has been already taken for single-phase codes, where at least three open-source codes for incompressible and fully compressible simulations are able to run on accelerated architectures: AFiD~\cite{zhu2018afid}, STREAmS~\cite{bernardini2021streams} and the accelerated version of CaNS~\cite{costa2021gpu}. Conversely, on the multiphase counterpart, despite the large availability of CPU-based open source codes, PARIS Simulator~\cite{aniszewski2021parallel}, TBFsolver~\cite{cifani2018highly}, FS3D~\cite{eisenschmidt2016direct}, NGA2~\cite{desjardins2008high}, Basilisk~\cite{popinet2009accurate} and MFC~\cite{Bryngelson_2020} to name few, limited effort has so far been devoted to their adaptation to hybrid architectures. \par
In this work, we aim to fill this gap and present FluTAS (Fluid Transport Accelerated Solver), a code for massive Direct Numerical Simulations on multi-GPU and multi-CPU architectures targeting incompressible multiphase flows, optionally with heat transfer. The numerical solution of these flows is typically performed using finite-difference methods in a staggered variable arrangement, and it involves the solution of a Poisson equation to enforce the constraints on the velocity divergence. In this context, FluTAS uses as basis the Navier-Stokes solver CaNS~\cite{costa2018fft} and its GPU extension~\cite{costa2021gpu}, whose key feature is a general implementation incorporating all the possible homogeneous pressure boundary conditions that can benefit of the FFT-based elliptic solvers~\cite{schumann1988fast}. The single-phase Navier-Stokes solver is extended to a two-fluid code using the one-domain formulation~\cite{scardovelli1999direct,prosperetti2009computational} and coupled with the algebraic VoF MTHINC~\cite{ii2012interface} to capture the two-phase interface. This method combines the exact mass conservation properties of certain geometric VoF methods with the reduced number of local operations for the interface reconstruction of the algebraic VoFs, making it a good candidate for properly exploiting hybrid and accelerated architectures. The version available in our group has been validated in~\cite{rosti2019numerical} and extensively employed in a different variety of multiphase configurations, both for laminar~\cite{de2019effect,de2020numerical,rosti2021shear} and turbulent~\cite{rosti2019droplets,kozul2020aerodynamically,crialesi2022modulation,cannon2021effect} flows. Note that it has been extended to phase changing flows~\cite{scapin2020volume} and also to handle weakly compressible multiphase flows (low-Mach approximation) \cite{dalla2021interface,scapin2022finite}. \par
%
%
This paper is organized as follows. In~\cref{sec:gov_eqn}, we introduce the governing equations for the incompressible two-fluid system. The discretization details of the VoF method, energy equation and Navier-Stokes solver are provided in \cref{sec:num_meth}, whereas the standard benchmarks for code validation are discussed in~\cref{sec:app_1}. Next, the parallelization for the GPU acceleration is presented together with the scaling tests in \cref{sec:gpu_parallel} and in \cref{sec:code_perf}. The code potentialities are shown in two demanding simulations of multiphase turbulence: emulsions in homogeneous isotropic turbulence (HIT) and two-phase thermal convection (see~\cref{sec:app_2}). Finally, main conclusions and future perspectives are summarized in~\cref{sec:concl}.

\section{Governing equations}\label{sec:gov_eqn}
We consider a two-phase system of immiscible incompressible Newtonian fluids (e.g., a gas-liquid system). The two phases are bounded by an infinitesimally small interface, through which momentum and energy can be transferred. To describe the system, we define a phase indicator function $H$ distinguishing the two phases at position $\mathbf{x}$ and time $t$:
\begin{equation}
	H(\mathbf{x},t) = \begin{cases}
						1 \hspace{0.5 cm} \text{if $\mathbf{x} \in \Omega_1$}\mathrm{,} \\
						0 \hspace{0.5 cm} \text{if $\mathbf{x} \in \Omega_2$}\mathrm{,} 
					  \end{cases}
	\label{ind_fun}
\end{equation}
where $\Omega_1$ and $\Omega_2$ are the domains pertaining to phases $1$ and $2$.  
We can use $H$ to define the thermophysical properties in the whole domain $\Omega = \Omega_1 \cup \Omega_2$ as follows:
\begin{equation}
    \xi(\mathbf{x},t) = \xi _1H(\mathbf{x},t)+\xi _2(1-H(\mathbf{x},t))\mathrm{,}
	\label{eqn:material_prop}
\end{equation}
where $\xi_{i}$ ($i=1,2$) can be the mass density $\rho_i$, the dynamic viscosity $\mu_i$, the thermal conductivity $k_i$ or the specific heat capacity at constant pressure $c_{p,i}$. Hereafter, unless otherwise stated, thermophysical quantities not specifically referring to one of the phases are defined from eq.~\eqref{eqn:material_prop}. The evolution of the indicator function is governed from the following topological equation:
\begin{equation}
  \dfrac{\partial H}{\partial t} + \nabla\cdot\left(\mathbf{u}_{\Gamma} H\right) =H\nabla\cdot\mathbf{u}_{\Gamma}\mathrm{,}
  \label{eqn:h_evol}
\end{equation}
where $\mathbf{u}_{\Gamma}$ is the interface velocity. In absence of phase change, the one-fluid velocity $\mathbf{u}$ is continuous across the interface and therefore, it can be employed as interface velocity in equation~\eqref{eqn:h_evol}. \par
The equations governing the momentum and energy transport for the liquid and gas phase are coupled through appropriate interfacial conditions~\cite{ishii2010thermo}, reported below in the so-called one-fluid or whole-domain formulation, where each transport equation is defined in $\Omega$ \cite{prosperetti2009computational}.

\begin{equation}
  \nabla\cdot{\mathbf{u}} = 0\mathrm{,}
  \label{eqn:kin_con}
\end{equation}

\begin{equation}
  \rho\left[\dfrac{\partial\mathbf{u}}{\partial t}+\nabla\cdot\left(\mathbf{u}\mathbf{u}\right)\right] = -\nabla p + \nabla\cdot\left[\mu\left(\nabla\mathbf{u}+\nabla\mathbf{u}^T\right)\right] + \sigma\kappa\delta_{\Gamma}+ \hat{\rho}\mathbf{g} \mathrm{,}
  \label{eqn:mom_con}
\end{equation}

\begin{equation}
  \rho c_p\left[\dfrac{\partial T}{\partial t}+\nabla\cdot\left(\mathbf{u}T\right)\right] = \nabla\cdot\left(k\nabla T\right)\mathrm{.} 
  \label{eqn:en_con}
\end{equation}
Here, $\mathbf{u}$ is the fluid velocity assumed to be continuous in $\Omega$, $p$ is the
hydrodynamic pressure, $T$ the temperature. In equation~\eqref{eqn:mom_con}, $\sigma$ is the surface tension, $\kappa$ the local interfacial curvature and $\delta_{\Gamma}$ is a delta Dirac function, $\mathbf{g}$ is the gravity acceleration and $\hat{\rho}$ is the volumetric density field modified to account for the thermal effects in the gravity forces. Using the Oberbeck–Boussinesq approximation, $\hat{\rho}$ reads as:
\begin{equation}
  \hat{\rho} = \rho_{1,r}\left[1-\beta_l\left(T-T_{r}\right)\right]H+\rho_{2,r}\left[1-\beta_g\left(T-T_{r}\right)\right](1-H)\mathrm{,}
\end{equation}
where $\rho_{i=1,2,r}$ are the reference phase densities and $\beta_{i=1,2}$ are the liquid and gas thermal expansion coefficients.

\section{Numerical methodology}\label{sec:num_meth} %
The numerical solution of the governing equations~\eqref{eqn:h_evol},~\eqref{eqn:kin_con},~\eqref{eqn:mom_con} and~\eqref{eqn:en_con} presented in section~\ref{sec:gov_eqn} is addressed on a fixed regular Cartesian grid with uniform spacing $\Delta x$, $\Delta y$ and $\Delta z$ along each direction. A marker-and-cell arrangement is employed for velocity and pressure points~\cite{harlow1965numerical}, whereas all scalar fields are defined at the cell centers. Each time-step, the governing equations are advanced in time by $\Delta t^{n+1}=t^{n+1}-t^{n}$, with the previous time-step indicated with $\Delta t^n=t^n-t^{n-1}$. Hereafter, we present the numerical discretization of the governing equations, following the same order in which they are solved.

\subsection{Volume of fluid: the MTHINC method}\label{subsec:mthinc}
The first step of the time-marching algorithm consists in the interface reconstruction and its subsequent advection. As previously mentioned, these tasks are addressed within a fully Eulerian framework using a volume-of-fluid (VoF) method. From a numerical point of view, this consists first in defining the volume fraction $\phi$ in each cell of the computational domain as:
\begin{equation}
  \phi=\dfrac{1}{V_c}\int_{V_c}H(\mathbf{x},t)dV_c\mathrm{,}
\end{equation}
with $V_c=\Delta x\Delta y\Delta z$. Next, equation~\eqref{eqn:h_evol} is written in terms of volume fraction as:
\begin{equation}
  \dfrac{\partial\phi}{\partial t}+\nabla\cdot\left(H\mathbf{u}\right)=\phi\nabla\cdot\mathbf{u}\mathrm{.}
\end{equation}
The distinct feature of each class of VoF method lies in the way $H$ is approximated. In this work we employ the algebraic volume-of-fluid method based on the Multi-dimensional Tangent Hyperbola reconstruction, MTHINC~\cite{ii2012interface}, whose central idea is to approximate $H$ with a hyperbolic tangent:
\begin{equation}
  H(\tilde{x},\tilde{y},\tilde{z}) = \frac{1}{2}\left[1+\tanh(\beta_{th}\left(\mathcal{T}(\tilde{\mathbf{x}})+d_{th}\right)\right]\mathrm{,}    
  \label{eqn:h_parameter}
\end{equation} 
\sloppy
where $\beta_{th}$, $d_{th}$ are the sharpness and the normalization parameter, respectively, and $(\tilde{x},\tilde{y},\tilde{z})$ a local coordinate system $\tilde{\mathbf{x}}=\left[\left(x-0.5\right)/\Delta x,\left(y-0.5\right)/\Delta y,\left(z-0.5\right)/\Delta z\right]$. Employing equation~\eqref{eqn:h_parameter} has two distinct advantages with respect to a piecewise approximation, commonly employed in the geometric VoF methods. First, the phase indicator $H$ can be approximated with a reconstructing polynomial $\mathcal{T}$ of arbitrary order in a straightforward manner. Next, once $\mathcal{T}$ is known, the resulting interface at the two-phase boundary has smooth but controlled thickness (with the parameter $\beta_{th}$), which also allows computing accurately the normal vector $\mathbf{n}$ and curvature tensor $\mathbf{K}$ directly from $\phi$. More details about the choice of $\mathcal{T}$ and the calculations of $d_{th}$, $\mathbf{n}$ and $\mathbf{K}$ are found in the original paper by Ii et al.~\cite{ii2012interface}, but for completeness we include them in the appendix~\ref{sec:details_mthinc} with the numerical implementation details. \\
After the reconstruction step, the interface is advected using a directional splitting approach~\citep{puckett1997high,aulisa2003geometrical}, which consists in evaluating the numerical fluxes sequentially in each direction using, for each split, the latest estimation of VoF field. Accordingly, three provisional fields $\phi_{i,j,k}^{p}$ (with $p=[x,y,z]$) are first computed:
\begin{equation}
  \phi_{i,j,k}^{p} = \dfrac{\phi_{i,j,k}^{s}-\dfrac{1}{\Delta l^p}\left[f_{+}^{p}(\phi_{i,j,k}^{s})-f_{-}^{p}(\phi_{i,j,k}^{s})\right]}{1-\dfrac{\Delta t^{n+1}}{\Delta l^p}\left(u_{+}^{p}-u_{-}^{p}\right)^n}\mathrm{,} 
  \label{eqn:phi_split}
\end{equation}
where $s=[n,x,y]$, $[\Delta l^x,\Delta l^y,\Delta l^z]=[\Delta x,\Delta y,\Delta z]$, $[u^x,u^y,u^z]=[u,v,w]$ with $u_{\pm}^p$ the $p$-th velocity component. The calculation of the numerical fluxes $f_{\pm}$ in equation~\eqref{eqn:phi_split} are evaluated using the hyperbolic tangent approximation of $H$ as detailed in appendix~\ref{sec:details_mthinc}. Next, the divergence correction step is applied in order to impose the volume conservation of both phases at a discrete level:
\begin{equation}
  \phi_{i,j,k}^{n+1} = \phi_{i,j,k}^z - \sum_{p=x,y,z} \dfrac{\Delta t^{n+1}}{\Delta l^p}\phi_{i,j,k}^p(u_+^p-u_-^p)^n\mathrm{.}
  \label{eqn:vof_corr_step}
\end{equation}
With the above approach, mass conservation is ensured up to the accuracy with which the divergence free condition~\eqref{eqn:kin_con} is satisfied. Accordingly, if direct methods are employed to solve the Poisson equation, the mass of each phase results to be conserved up to machine precision. Another approach with a similar property has been introduced in~\cite{weymouth2010conservative}. However, in that case the dilatation terms at the denominator of equation~\eqref{eqn:phi_split} are treated in an explicit manner, while here in an implicit strategy is employed. This comes at a cost of the final correction step, given by equation~\eqref{eqn:vof_corr_step}, but with the advantage of not introducing additional time-step restrictions (apart the convective one) in the advection of the color function. 

\subsection{Thermal effects}\label{subsec:energy}
The next step of the time-marching algorithm consists in advancing the temperature field using an explicit second-order Adams-Bashforth method: 
\begin{equation}
  T^{n+1} = T^n+\Delta t^{n+1}\left(f_{t,1}\mathcal{M}_T^n-f_{t,2}\mathcal{M}_T^{n-1}\right)\mathrm{,}
  \label{eqn:tmp}
\end{equation}
where $f_{t,1}=(1+0.5\Delta t^{n+1}/\Delta t^n)$ and $f_{t,2}=0.5\Delta t^{n+1}/\Delta t^n$ are the coefficients of the Adams-Bashforth scheme. In equation~\eqref{eqn:tmp}, the operator $\mathcal{M}_T$ accounts for the advection and diffusion contribution and it is provided below in a semi-discrete form:
\begin{equation}
  \mathcal{M}_T^n = -\nabla\cdot(\mathbf{u}^nT^n) + \dfrac{1}{\rho^{n+1}c_p^{n+1}}\nabla\cdot(k^{n+1}\nabla T^n)\mathrm{.}
  \label{eqn:rt}
\end{equation}
All the spatial terms in equation~\eqref{eqn:rt} are discretized with second-order central schemes, except for the temperature convection term. The discretization of the latter is based on the 5th-order WENO5 scheme, as in reference~\cite{castro2011high}.

\subsection{Pressure correction algorithm}\label{subsec:pc}
Once the energy equation has been advanced, the momentum equation is solved with a second-order pressure correction~\cite{chorin1968numerical}, reported below in a semi-discrete form:
\begin{align}
   \left(\dfrac{\mathbf{u}^{\star\star}-\mathbf{u}^{n}}{\Delta t^{n+1}}\right) = f_{t,1}\mathcal{M}_\mathbf{u}^{n}-f_{t,2}\mathcal{M}_\mathbf{u}^{n-1}+\dfrac{\left(\sigma\kappa\delta_{\Gamma}+\hat{\rho}\mathbf{g}\right)^{n+1}}{\rho^{n+1}}\mathrm{,}\label{eqn:uss_predic} \\
   \mathbf{u}^{\star}=\mathbf{u}^{\star\star}-\dfrac{\Delta t^{n+1}}{\rho_0}\left[\left(1-\dfrac{\rho_0}{\rho^{n+1}}\nabla\hat{p}\right)+\nabla p^{n}\right]\mathrm{,} \label{eqn:us_predic} \\
   \nabla^2\psi^{n+1} = \dfrac{\rho_0}{\Delta t^{n+1}}\nabla\cdot\mathbf{u}^{\star}\label{eqn:poisson}\mathrm{,} \\
   \mathbf{u}^{n+1}=\mathbf{u}^{\star}-\dfrac{\Delta t^{n+1}}{\rho_0}\nabla \psi^{n+1}\label{eqn:corr}\mathrm{,} \\
  p^{n+1}=p^n+\psi^{n+1}\mathrm{,}
  \label{eqn:p_upt}
\end{align}
where the operator $\mathcal{M}_\mathbf{u}^n$ and $\mathcal{M}_\mathbf{u}^{n-1}$ in equation~\eqref{eqn:uss_predic} includes the convective and diffusive terms computed at the current and previous time level, neglecting the surface tension and gravity forces which are then included as source terms. The spatial gradients in $\mathcal{M}_\mathbf{u}$ are discretized with central schemes. The intermediate velocity $\mathbf{u}^{\star\star}$ is then updated with the contribution from the terms due to the time-pressure splitting, as in~\eqref{eqn:us_predic}. Note that $\rho_0$ is the minimum value of the density field in the computational domain and $\hat{p}$ represents the time-extrapolated pressure between the current and the old time step, i.e. $\hat{p}=(1+\Delta t^{n+1}/\Delta t^n)p^n-(\Delta t^{n+1}/\Delta t^n)p^{n-1}$. Following~\cite{frantzis2019efficient} and contrary to~\cite{dong2012time,dodd2014fast}, the terms arising from the pressure splittings are included in the prediction of the velocity field (see eq.~\eqref{eqn:us_predic} before the imposition of the boundary conditions. This approach has two distinct advantages. First, it represents an incremental pressure projection which allows achieving an almost second-order accurate in time pressure field~\cite{frantzis2019efficient}. Next, it ensures the consistency of the pressure field near a solid boundary (i.e, $\mathbf{u}^{n+1}=\mathbf{u}^{\star}=0$), where the pressure gradient component normal to the boundary (i.e., $\nabla_{\perp}\psi^{n+1}=0$) vanishes independently of the local density (see eq.~\eqref{eqn:corr}). \par
Next, the constant coefficients Poisson equation~\eqref{eqn:poisson} is solved with the method of eigenexpansion technique that can be employed for different combination of homogeneous pressure boundary conditions~\cite{schumann1988fast}. Finally, the velocity field is corrected as in equation~\eqref{eqn:corr} in order to impose the divergence constrain (i.e., solenoidal velocity field) and the pressure updated as in equation~\eqref{eqn:p_upt}.

\subsubsection{Poisson solver}
The code uses the FFT-based finite-difference direct solver developed and implemented in the DNS code CaNS; see \cite{costa2018fft,costa2021gpu}. The underlying numerical approach dates back to the late 1970s \cite{swarztrauber1977methods,schumann1988fast}, and has regained popularity in recent years, thanks to the improvements of hardware, and of and software frameworks for collective data communications, provided by the MPI standard and higher-level libraries like 2DECOMP\&FFT. In a nutshell, the approach uses Fourier-based expansions along two domain directions, which reduce the system of equations resulting from the three-dimensional second-order finite-difference Laplace operator (seven non-zero diagonals) to a simple, tridiagonal system. These Fourier-based expansions depend on the boundary conditions of the system, and can be computed using FFTs, some of them with pre-/post-processing of the FFT input/output (see, e.g., \cite{makhoul1980fast}). \\
The FFT-based expansions are employed along directions $x$ and $y$, and the resulting tridiagonal system along $z$ is then solved using Gauss elimination. For calculations on CPUs, the method leverages the \emph{guru} interface of the FFTW library \cite{frigo1998fftw}, which allows for performing all possible combinations of discrete transforms using the same syntax. On GPUs, the fast discrete cosine and sine transforms have been implemented using real-to-complex/complex-to-real FFTs from the cuFFT library, with pre- and post-processing of the input and output signals to calculate the desired series expansion \cite{makhoul1980fast,costa2021gpu}. We refer to Refs.~\cite{costa2018fft,costa2021gpu} for more details on this method and its implementation. \\
Concerning the parallelization of the method in a distributed-memory setting, the FFT-based transforms and Gauss elimination steps require the data along each direction to be local to each MPI task. The domain is decomposed using a 2D pencil decomposition, where collective \emph{all-to-all} communications are required to transpose the orientation of the 2D data decomposition. These transposes are performed using the 2DECOMP\&FFT library \cite{li20102decomp}, which was modified to allow for GPU-GPU communication in \cite{zhu2018afid,costa2021gpu}. \\
It is worth noting that, in line with the recent developments of CaNS, the present method uses a default decomposition (i.e., ``outside'' the Poisson solver) based on a partitioning along $y$ and $z$, resulting in $x$-aligned pencils. This reduces the total number of data transposes to be performed during the solution of the Poisson equation from $6$ to $4$. The approach has been adopted for both CPUs and GPUs, and the required operations to solve the Poisson equation are summarized as follows:
\begin{enumerate}
  \item perform forward FFT-based transforms along $x$;
  \item transpose $x$-to-$y$;
  \item perform forward FFT-based transforms along $y$;
  \item transpose $y$-to-$z$;
  \item solve tridiagonal system using Gauss elimination along $z$;
  \item transpose $z$-to-$y$;
  \item perform backward FFT-based transforms along $y$;
  \item transpose $y$-to-$x$;
  \item perform backward FFT-based transforms along $x$.
\end{enumerate}
Moreover, for the GPU implementation, the solver explicitly reduces the number of \emph{all-to-all} operations when the domain is not decomposed along $z$ (i.e., when a $x-y$ slab decomposition is prescribed). This effectively decreases the number of collective operations from $4$ to $2$ (steps $2$ and $8$ above are skipped). This is the approach adopted in the GPU runs presented here -- due to the higher memory bandwidth in GPUs, a slab decomposition suffices for distributed-memory calculations with sufficiently small wall-clock time per step. Explicitly skipping these two no-op resulted in a substantial reduction in wall-clock time per step, and in an overall improvement in the parallel scalability of the solver.

\subsection{Complete solution algorithm}
For clarity, a step by step description of the overall solution procedure is presented in Algorithm~\ref{tab:algorithm}.
\begin{algorithm}
\caption{Overall solution procedure}
\begin{algorithmic}[1]
\State $\phi^0, \ T^0, \ \mathbf{u}^0, \ p^0$ are initialized;
\State $\rho^0$, $\mu^0$, $k^0$ and $c_p^0$ are calculated using equation~\eqref{eqn:material_prop} from $\phi^0$;
\State $n=0$ is set,
\While{($t<t_{tot}$ $\parallel$ $n<N_{tot}$)}
    \State Set $n=n+1$ and $\Delta t^{n+1}$;
    \State $\phi^{n+1}$ is calculated from equation~\eqref{eqn:phi_split} and~\eqref{eqn:vof_corr_step};
    \State $\mathbf{n}^{n+1}$ and $\kappa^{n+1}$ are evaluated using the procedure described in~\ref{sec:details_mthinc};
    \State $\rho^{n+1}$, $\mu^{n+1}$, $k^{n+1}$ and $c_p^{n+1}$ is calculated from equation~\eqref{eqn:material_prop};
    \State $T^{n+1}$ is calculated from Eq.~\eqref{eqn:tmp};
    \State $\mathbf{u}^{*}$ is calculated from Eq.~\eqref{eqn:uss_predic} and Eq.~\eqref{eqn:us_predic};
    \State $\psi^{n+1}$ is calculated from Eq.~\eqref{eqn:poisson};
    \State $\mathbf{u}^{n+1}$ is calculated from Eq.~\eqref{eqn:corr};
    \State $p^{n+1}$ is computed from Eq.~\eqref{eqn:p_upt}.
\EndWhile
\State End of simulation.
\end{algorithmic}
\label{tab:algorithm}
\end{algorithm}

\section{Validation}\label{sec:app_1} 
\subsection{Two-dimensional Zalesak's disk}\label{sec:zal} 
The Zalesak problem represents a classical benchmark to assess the accuracy of the interface capturing/tracking algorithm. It consists in the solid-body rotation of a slotted disk immersed in an imposed two-dimensional velocity field $\mathbf{u}=(0.5-y,x-0.5)$. The disk can be easily defined in a Cartesian two-dimensional domain by setting the indicator function $H_{i,j,k}^0$ equal to $1$ inside the following domain $\Omega_{H}$
\begin{equation}
  \Omega_H: \left[\left(x-0.5\right)^2+\left(y-0.75\right)^2\right] \leq 0.15^2 \cap \left(|x-0.5|\geq 0.025 \cup y\geq 0.85\right)\mathrm{.}
\end{equation}
The benchmark consists in comparing the deformation of the solid disk with respect to the initial shape after one entire revolution. The VoF equation is solved in a two-dimensional square domain $\Omega=[0,1]\times[0, 1]$, discretized with four different grid spacing $[\Delta x,\Delta y]=[1/N_x,1/N_y]$ with $N_x\times N_y=[32\times 32, 64\times 64, 128\times 128, 256\times 256]$. Periodic boundary conditions are prescribed in both directions. Simulations are conducted up to $t=2\pi$ (i.e., one complete revolution of the slotted disk) using a constant time-step $\Delta t=t/3200$. Note that this value has been chosen to ensure a stable time integration for the highest grid resolutions cases (i.e., $256\times 256$) and is employed for the coarser cases. 
\begin{figure}[h!]
    \centering
    \includegraphics[width=\textwidth]{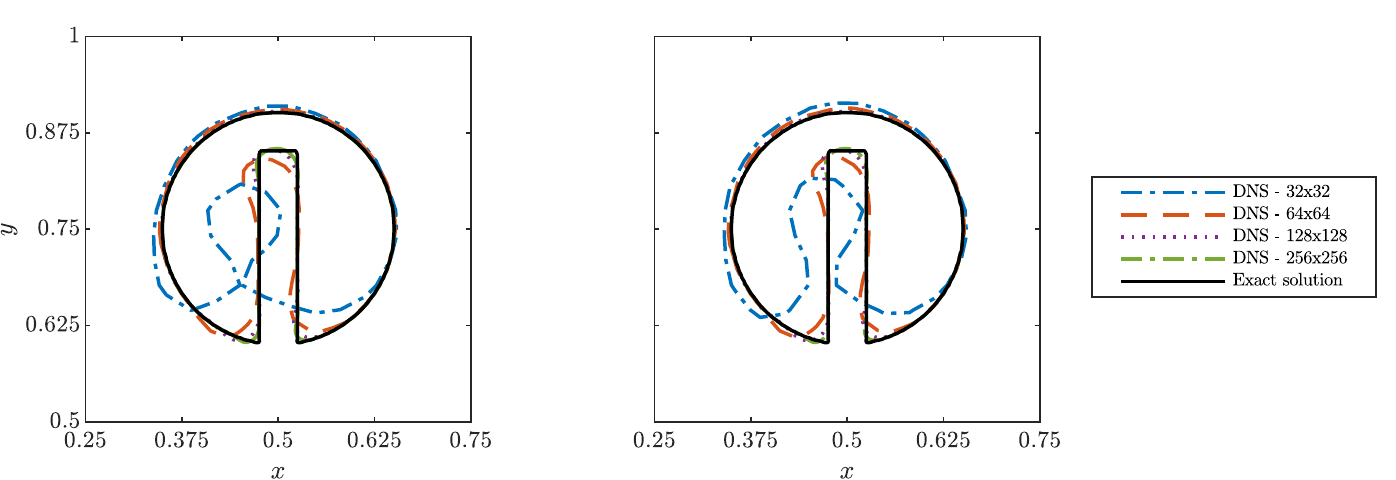}
    \caption{Deformation of the Zalesak's slotted disk after $t=2\pi$ for $\beta_{th}=2$ (left) and $\beta_{th}=3$ (right).}
	\label{fig:zal_disk_con}
\end{figure}%
Figure~\ref{fig:zal_disk_con} shows the final disk shape for different grid solutions and for two sharpness parameters $\beta_{th}=2$ and $\beta_{th}=3$. Note that the highest deviation from the initial shape are in the corner regions, where the high-curvature regions are located. Moreover, the solution is weakly dependent on the value of $\beta_{th}$ and deviations between the different employed $\beta_{th}$ are visible only for the coarser simulations. \\
Finally, to assess the accuracy of the solution, we compute the $L_1$ norm and the order of convergence as:
\begin{equation}
  L_1 = \dfrac{1}{N_xN_y}\sum_{i=1}^{N_x}\sum_{y=1}^{N_z}|\phi(i,j)-\phi^0(i,j)|\mathrm{,}
\end{equation}
\begin{equation}
  n_{L1} = \dfrac{\log\left(\dfrac{L_{2,N}}{L_{1,N}}\right)}{\log(2)}\mathrm{,}
\end{equation}
where $L_{1,N}$ is the $L_1$-error using $N_x\times N_y$ grid points and $L_{1,2N}$ is the $L_1$-error
evaluated with $2N_x\times 2N_y$ grid points.
\begin{figure}[h!]
    \centering
    \includegraphics[width=0.5\textwidth]{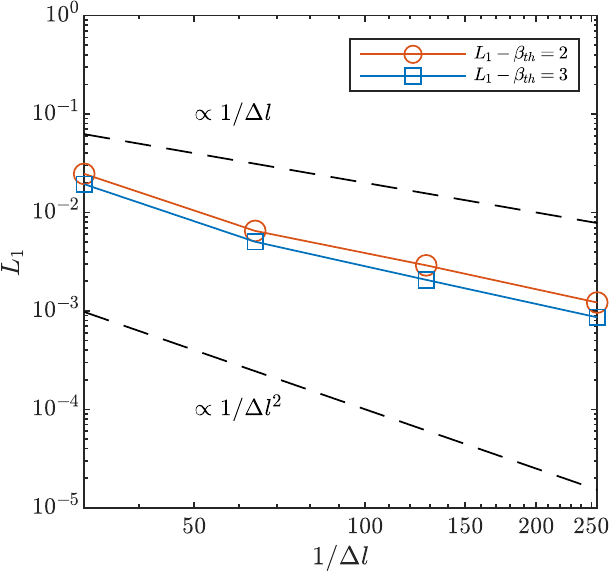}
    \caption{Deformation of the Zalesak's slotted disk after $t=2\pi$ for $\beta_{th}=2$ (left) and $\beta_{th}=3$ (right).}
	\label{fig:zal_disk_err}%
\end{figure}
Results are reported in figure~\ref{fig:zal_disk_err}, where an order of convergence between the first and the second-order is achieved for $\phi$, almost independent of the employed value of $\beta_{th}$.

\subsection{Three-dimensional rising bubble}\label{sec:3d_ris} 
The rising bubble test case is a well-established numerical benchmark for multiphase flows~\cite{turek2017numerical}. This test is presented here to showcase the ability of the numerical tool to accurately capture the topological changes of a moving interface. The flow is driven by the density difference between the two phases, and is influenced by the viscosity difference and the surface tension. The relevant dimensionless groups for this flow are the Reynolds number Re$=\rho_{g}u_rl_r/\mu_{g}$, the Weber number We$=\rho_{g}u_{r}^2l_r/\sigma$, the Froude number Fr$=u_r/\sqrt{|\mathbf{g}|d_0}$, the density ratio $\lambda_{\rho}=\rho_{l}/\rho_{g}$ and the viscosity ratio $\lambda_{\mu}=\mu_{l}/\mu_{g}$. In these definitions, $l_r$ is a reference length and $u_r$ the reference velocity. Moreover, $\sigma$ is the surface tension coefficient, $\rho_{g}$ and $\rho_{l}$ the reference gas and liquid densities, and $\mu_{g}$ and $\mu_{l}$ the reference gas and liquid dynamic viscosity. Finally, $\mathbf{g}$ is the acceleration of gravity and $d_0$ is the initial diameter of the spherical bubble. \\
Following the benchmark study~\cite{turek2017numerical}, the values adopted for the dimensionless groups are Re$=35$, We$=1$, Fr$=1$, $\lambda_{\rho}=10$ and $\lambda_{\mu}=10$, setting $l_r=d_0$, $u_r=\sqrt{|\mathbf{g}|d_0}$ and the reference time $t_r=\sqrt{d_0/|\mathbf{g}|}$. The dimensions of the computational domain are $l_x=l_y=2d_0$ and $l_z=4d_0$. The acceleration of gravity acts along the $z$-direction. No-slip and no-penetration boundary conditions are prescribed at the horizontal top and bottom boundaries of the domain ($z$-normal) and periodic conditions are prescribed at the vertical boundaries ($x$- or $y$-normal). A uniform Cartesian grid of 128$\times$128$\times$256 cells is used. Initially, stagnant flow conditions are applied and the position of the center of mass of the spherical bubble, denoted as $(x_{c}(t),y_{c}(t),z_{c}(t))$, is located at $(d_0,d_0,d_0)$. A constant time-step $\Delta t / t_r=2.8\times10^{-4}$ is used to advance the solution in time. 
%
\begin{figure}[t!]
    \centering
    \includegraphics[width=1.0\textwidth]{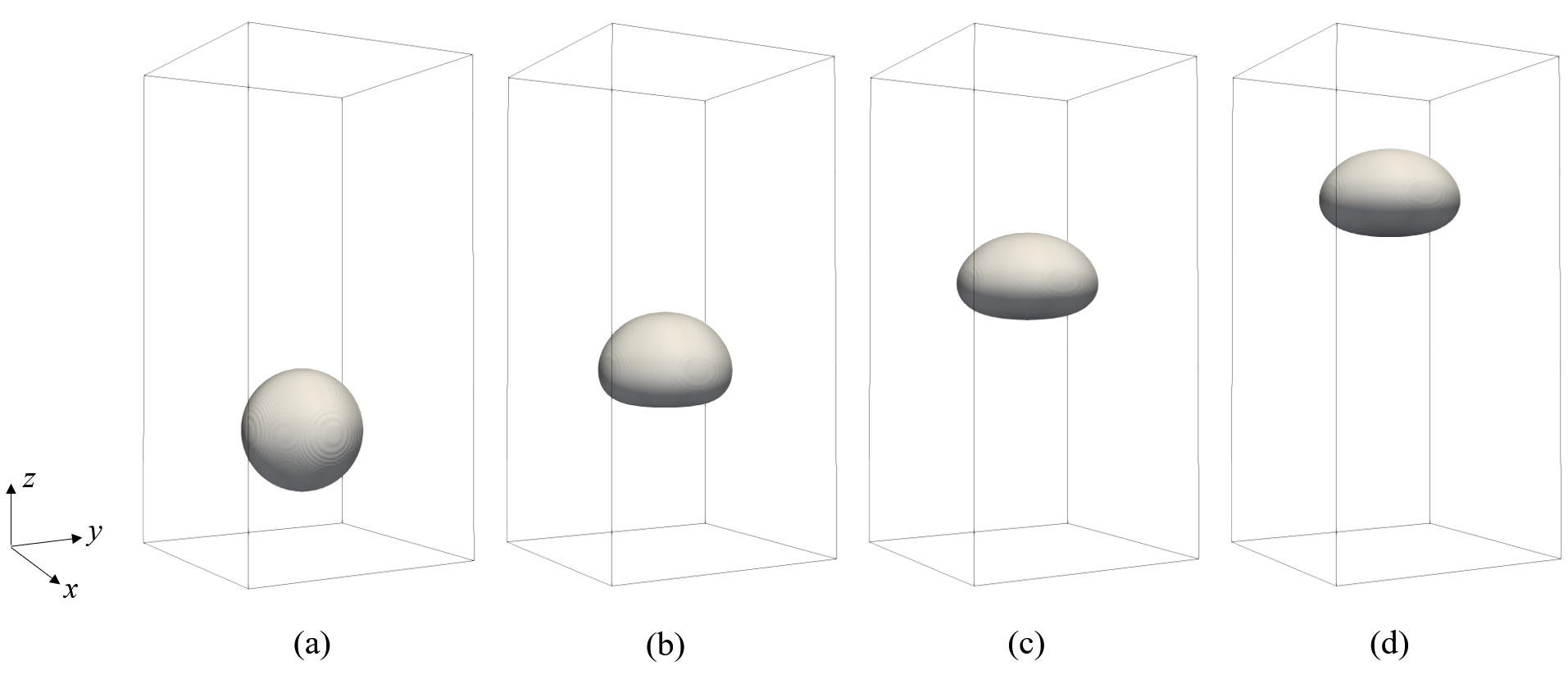}
    \caption{Isosurfaces of $\phi=0.5$ at dimensionless times $t \sqrt{|\mathbf{g}|/d_0}$ (a) 0, (b) 1.4, (c) 2.8 and (d) 4.2.}
	\label{fig:rising_bubble1}%
\end{figure}
%
\begin{figure}[t!]
    \centering
    \includegraphics[width=0.6\textwidth]{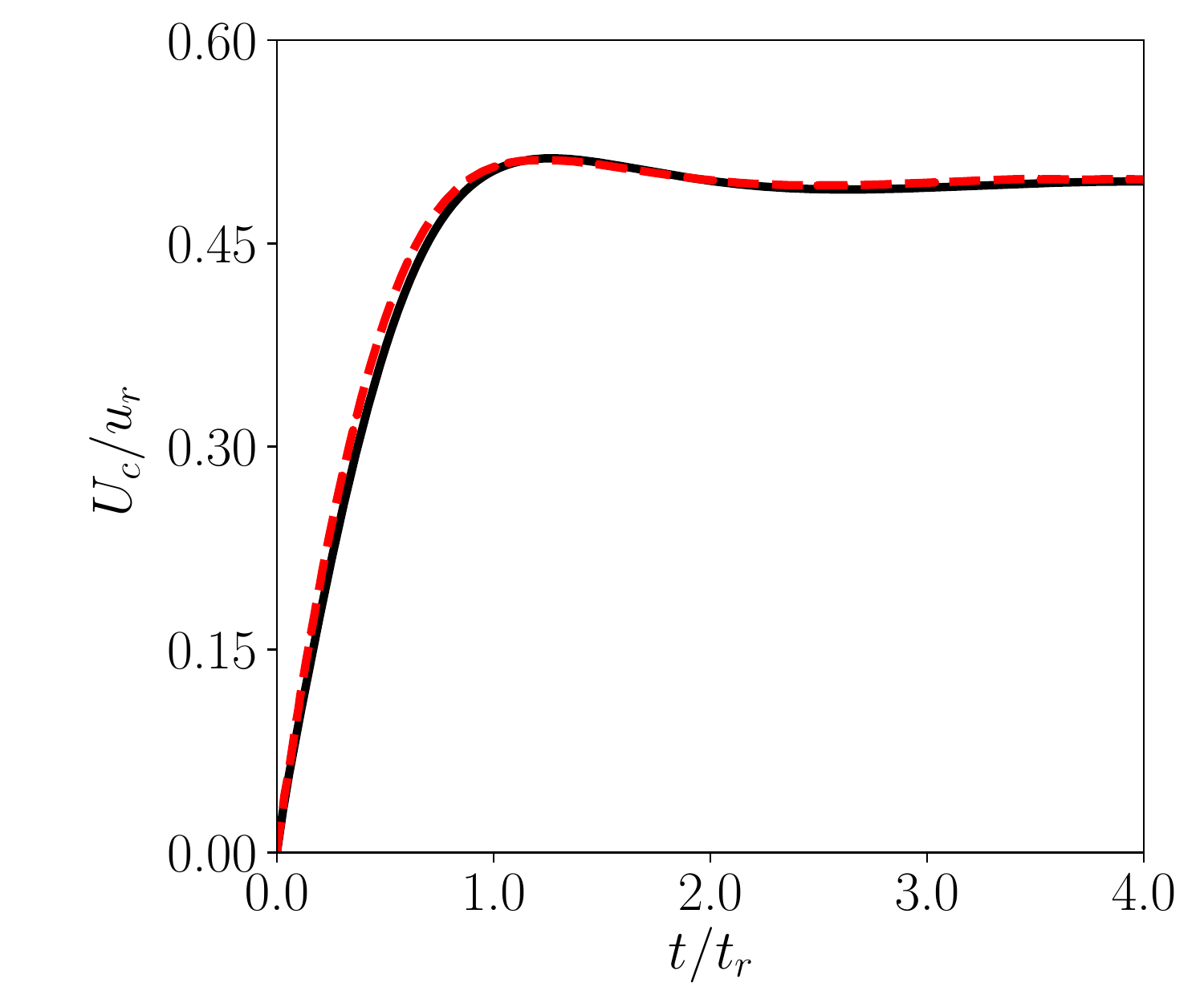}
    \caption{Bubble rise velocity as a function of time. Black solid line, present results; red dashed line, reference results from~\cite{turek2017numerical}.}
	\label{fig:rising_bubble2}%
\end{figure}
%
Figure~\ref{fig:rising_bubble1} shows the isosurfaces of $\phi=0.5$ at various time instances. It is evident that as the initially spherical bubble rises, its surface topology changes. To compare against the reference results from~\cite{turek2017numerical}, figure~\ref{fig:rising_bubble2} shows the evolution of the bubble rising velocity $U_c$ in time. The bubble velocity is defined as,
\begin{equation}
  U_c=\frac{\displaystyle{\int_{\Omega} \phi \ w \ d\Omega}}{\displaystyle{\int_{\Omega} \phi \ d\Omega}}\mathrm{,}
\end{equation}
\begin{table}[t]
\centering
\begin{tabular}{lccc}
\hline
 & $t / t_r$ & $U_c/u_r$ & $A_0/A$ \\
\hline
ref.~\cite{turek2017numerical} &  & 0.51013 & 0.97418 \\
present & 1.4 & 0.51144 & 0.97892 \\
dev. \% & & 0.26 & 0.49 \\
\hline
ref.~\cite{turek2017numerical} &  & 0.49823 & 0.95925\\
present & 4.2 & 0.49512 & 0.96057 \\
dev. \% & & 0.62 & 0.14\\
\hline
\end{tabular}
\caption{Comparison of the dimensionless rise velocity $U_c/u_r$ and bubble sphericity $A_0/A$ between reference and present results.}
\label{table:rising_bubble}
\end{table}
where $w$ is the vertical velocity component and $\Omega$ the volume of the entire domain. After an initial period where the bubble accelerates, the rise velocity reaches a maximum and then stabilizes. Figure~\ref{fig:rising_bubble2} demonstrates an excellent agreement between present and reference results. To further quantify this agreement, table~\ref{table:rising_bubble} presents benchmark quantities for comparison at specific time instances. Besides the bubble velocity, the table shows the bubble sphericity $A_0/A$, defined as the initial value of the bubble surface area over the value at a later time. The deviation between reference and present values is less than 1\%.  

\subsection{Differentially heated cavity}\label{sec:dhc} 
To demonstrate the accuracy of the code in the presence of thermal effects, this section considers the flow of air in a closed two-dimensional square heated cavity. The cavity is heated and cooled by the vertical side walls (y-normal), while the horizontal walls are adiabatic (z-normal). Within this configuration, a circulation is formed and maintained by the ascending hot fluid next to the heated wall and the descending cold fluid next to the cooled wall.
\begin{figure}[h!]
    \centering
    \includegraphics[width=7.00 cm, height=5.25 cm]{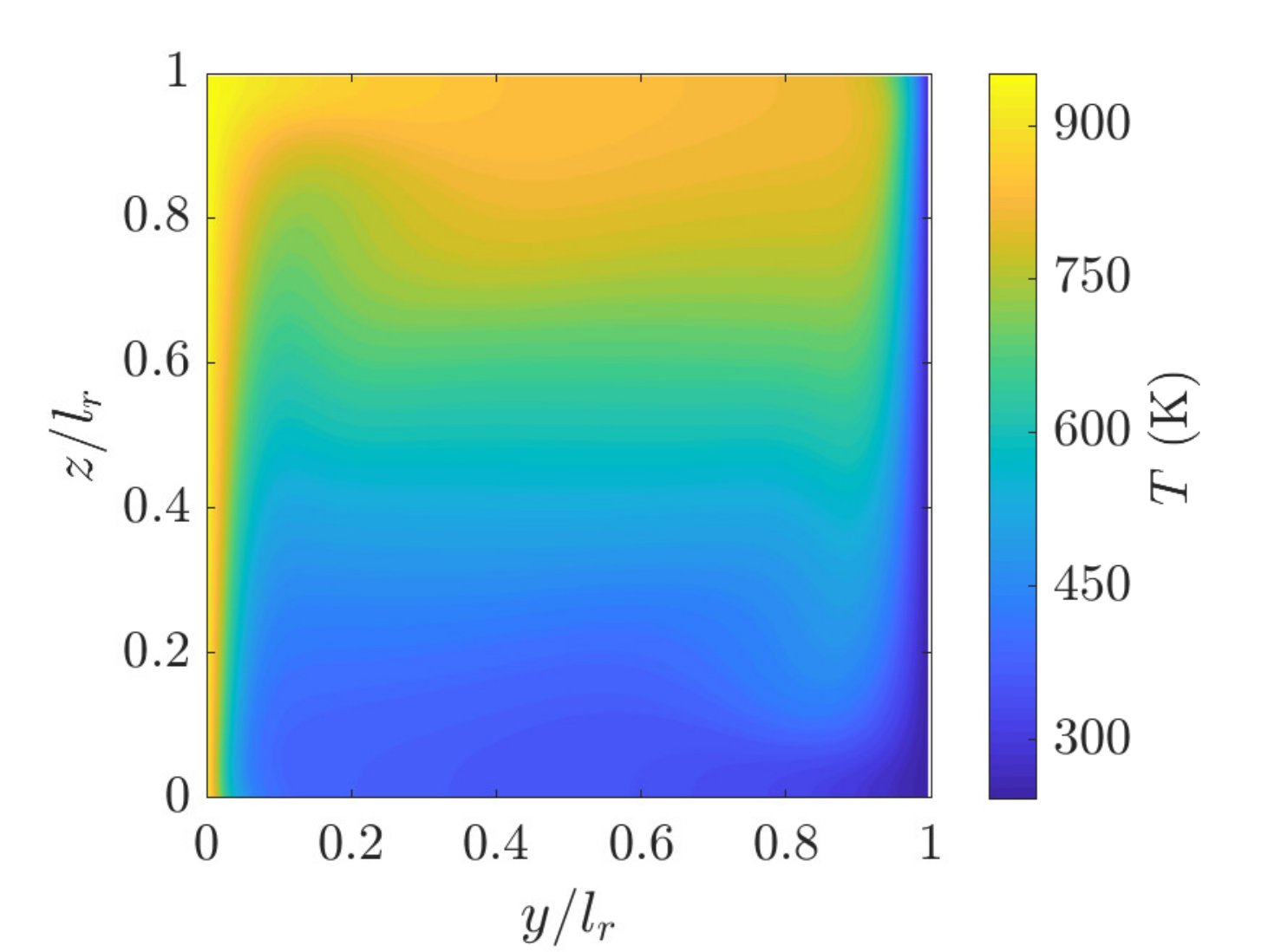}
    \hspace{-0.5 cm}
    \includegraphics[width=6.50 cm, height=5.00 cm]{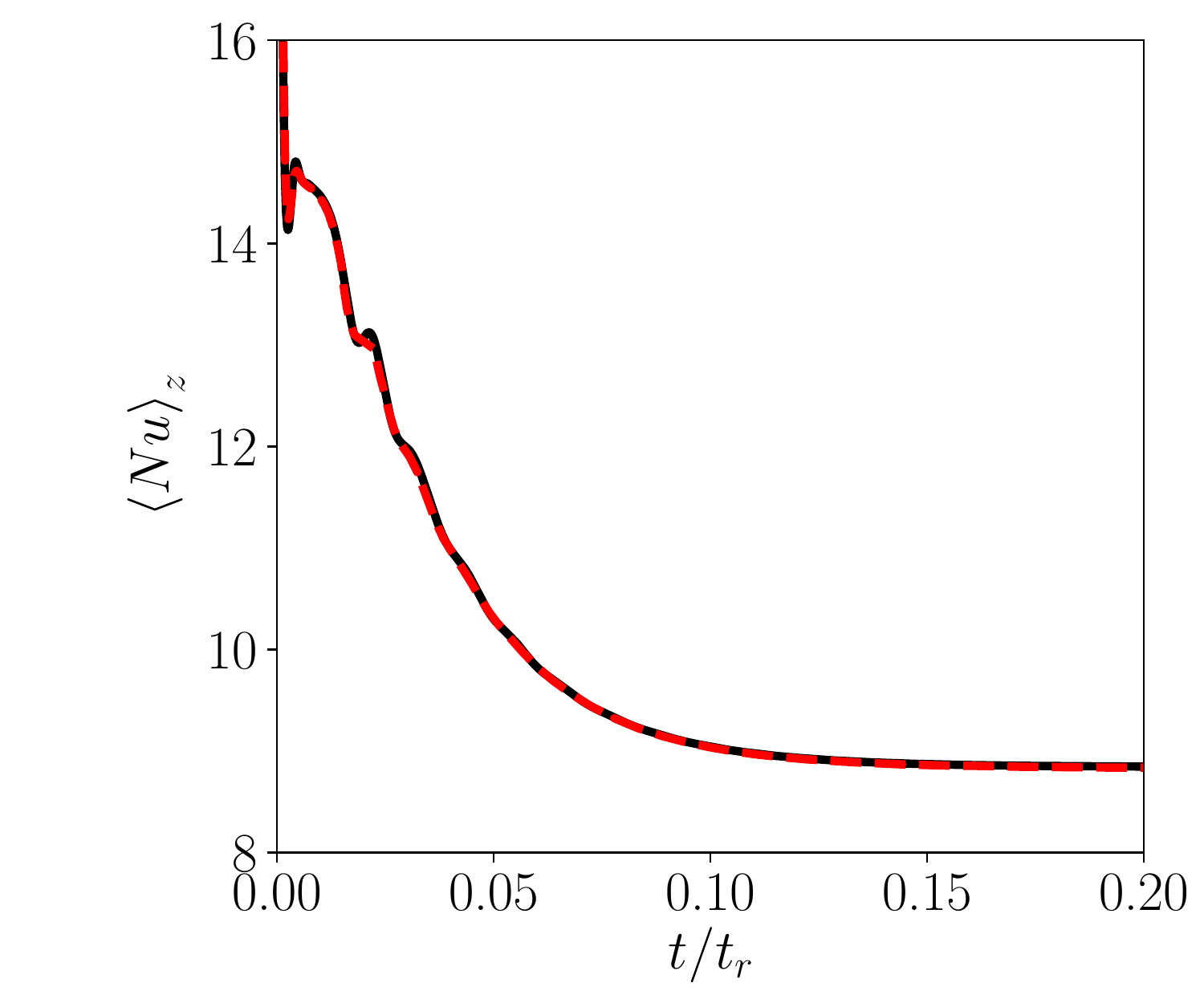}
    \put(-380,125){(\textit{a})}
    \put(-180,125){(\textit{b})}
    \caption{(\textit{a}) Contour plot of the temperature field at $t/t_r=0.5$ (steady state) for the differentially heated cavity test case, (\textit{b}) Temporal evolution of the wall-averaged Nusselt number on the heated wall. Black solid line, present results; red dashed line, reference results from~\cite{armengol2017effects}.}
	\label{fig:dhc1_dhc2}
\end{figure}
%
%
The flow is therefore purely thermally-driven and is characterized by the Rayleigh number Ra$=|\mathbf{g}|\beta\Delta T l_r^3 / \left(\nu \alpha\right)$ and the Prandtl number Pr$=\nu / \alpha$. In these definitions, $\beta$ is the fluid thermal expansion coefficient, $\nu$ is the fluid viscosity, $\alpha$ is the fluid thermal diffusivity and $\Delta T=(T_h-T_c)$ is the temperature difference between the heated ($T_h$) and cooled ($T_c$) walls. Typically, the height of the cavity is taken as the reference length ($l_r=L_z$), while the reference velocity and time are defined as $u_r=\alpha/l_r$ and $t_r=l_r^2/\alpha$. The case simulated here follows the setup presented in several studies~\cite{de1983natural,leal2000integral,armengol2017effects} with Ra$=10^6$ and Pr$=0.71$. The domain boundaries are  solid walls, and no-slip boundary conditions are applied. With respect to the temperature field, constant temperature boundary conditions are applied on the vertical walls and a zero temperature gradient along the normal direction is applied on the horizontal walls. The domain is discretized in space using a uniform Cartesian grid with $256\times256$ cells.
Initially, the air in the cavity is stagnant and isothermal at a temperature $T_0=T_c$. A constant time-step $\Delta t$ is used to advance the solution in time, given by $\Delta t/t_r=5.0\times10^{-7}$. Figure~\ref{fig:dhc1_dhc2}\textit{a}) shows the contour of the temperature field at $t/t_r=0.5$, at which point a steady state has been reached. The temperature field is characterized by thin and spatially developing thermal boundary layers next to the thermally active vertical walls, and a stratified region at the central area of the cavity. The heat transfer rate inside the cavity is expressed through the Nusselt number, defined as,
\begin{equation}
  \mathrm{Nu}=\frac{h l_r}{k}=\frac{l_r}{\Delta T}\mathbf{\nabla}T\big|_w\cdot\mathbf{n}_w\mathrm{,}
\end{equation}
where $h$ is the heat transfer coefficient, $k$ is the fluid thermal conductivity, $\mathbf{\nabla}T\big|_w$ is the temperature gradient on any of the thermally active vertical walls and $\mathbf{n}_w$ is the corresponding unit normal vector on the wall. Figure~\ref{fig:dhc1_dhc2}\textit{b}) shows the comparison of the temporal evolution of the wall-averaged Nusselt number $\left<Nu\right>_z$ on the heated wall between the present and reference results from~\cite{armengol2017effects}. It is evident that the present results are in excellent agreement with the reference solution for the entire duration of the simulation. Furthermore, Table~\ref{table:dhc} presents the comparison of key benchmark quantities at steady state, confirming the agreement between present and reference results. 

\begin{table}[t!]
\centering
\begin{tabular}{lccccc}
\hline
 & $V_{max}/u_r$ & $W_{max}/u_r$ & $Nu_{max}$ & $Nu_{min}$ & $\left<Nu\right>_{z}$\\
\hline
Ref.~\cite{armengol2017effects} & 64.85 & 220.6 & 17.58 & 0.9794 & 8.830 \\
Present & 64.86 & 220.3 & 17.67 & 0.9773  & 8.843 \\
\% dev. & 0.02 & 0.14 & 0.51 & 0.21 & 0.14 \\
\hline
\end{tabular}
\caption{Comparison of key benchmark quantities at steady state for the differentially heated cavity test case. $V_{max}$ is the maximum horizontal velocity along the vertical mid-plane ($y=0.5l_r$), $W_{max}$ is the maximum vertical velocity along the horizontal mid-plane ($z=0.5l_r$), $Nu_{max}$ and $Nu_{min}$ are the maximum and minimum values of the Nusselt number on the heated wall, and $\left<Nu\right>_{z}$ is the averaged Nusselt number value on the heated wall.}
\label{table:dhc}
\end{table}

\section{Code parallelization and GPU acceleration}\label{sec:gpu_parallel} 
 \subsection{Domain decomposition}
The code is designed to run both in multi-CPU and multi-GPU architectures. For domain decomposition, both slab (1D) and pencil (2D) are allowed \citep{costa2018fft} through the library 2DECOMP \citep{li20102decomp}. The type of decomposition can be implicitly set via the 2-component array \code{dims} (e.g. $[1,n]$ for slabs and $[n,m]$ for pencils) in one input file \code{dns.in}. The pencil/slab orientation can be arbitrarily chosen as in CaNS, via the preprocessor flags \code{-D\_DECOMP\_X}, \code{-D\_DECOMP\_Y} and \code{-D\_DECOMP\_Z} which set the direction over which the domain is not decomposed. This flexibility allows improving the efficiency of both the CPU and the GPU implementations. For CPU, using pencils allows increasing the number of processes used per execution (i.e. up to $N^2$ for $n_x=n_y=n_z=N$), hence reducing the time to solution. In the GPU implementation, only the \textit{z-pencil} and \textit{x-slabs} decompositions are allowed. It is recommendable to use \textit{x-slabs} on GPU (i.e. compiling with \code{-D\_DECOMP\_X} and using \code{dims}$=[1,n]$) as this implementation reduces the number of \emph{all-to-all} calls to the minimum, hence reducing GPU-GPU communication and improving performances on multi-nodes runs.

\subsection{Code parallelization}
The parallelization is performed using MPI. When GPU acceleration is enabled, MPI allocates one rank for each GPU. The code assumes the chosen MPI library is “CUDA-aware”, meaning GPU data is directly passed to MPI function call and the MPI implementation takes care of moving the data in the most efficient way. If available, GPU-to-GPU communication can leverage NVIDIA NVLink which is a physical GPU-to-GPU interconnection known to have higher bandwidth (at least one order of magnitude) than Infiniband. Throughout the code, all nested for-loops, i.e. iterations over all the domain points, are accelerated on GPUs using OpenACC \cite{openaccW}, a portable standard directive-based programming model that can execute code on multi-core CPUs as well as accelerators like NVIDIA GPU. Such offload is not used for CPU-only compilation and execution.
To execute FluTAS, the platform needs to support NVIDIA Unified Memory, which has two main advantages:
\begin{itemize}
	\item the ability of allocating and managing more GPU allocated memory than what is physically present on the device;
	\item the ability to avoid explicitly to handle data movements Host-to-Device and Device-to-Host, leaving the runtime do the work for the developers.
\end{itemize}
Both features are used in the code and proved crucial for an efficient GPU acceleration.


\section{Code performance}\label{sec:code_perf} 
We now present an analysis of the code performances on standard CPU-based  and accelerated GPU-based architectures. Tests on GPUs were performed on MeluXina at \textit{LuxProvide} (LXP, Luxembourg) \cite{meluxinaW} and Berzelius at \textit{National Supercomputer Centre} (NSC, Sweden) \cite{berzeliusW}, while tests on CPUs  were performed on Tetralith also managed by NSC. 
\subsection{Weak and strong scaling}\label{subsec:weak_scaling}
\begin{figure}[h!]
	\centering
	\includegraphics[width=0.49\textwidth]{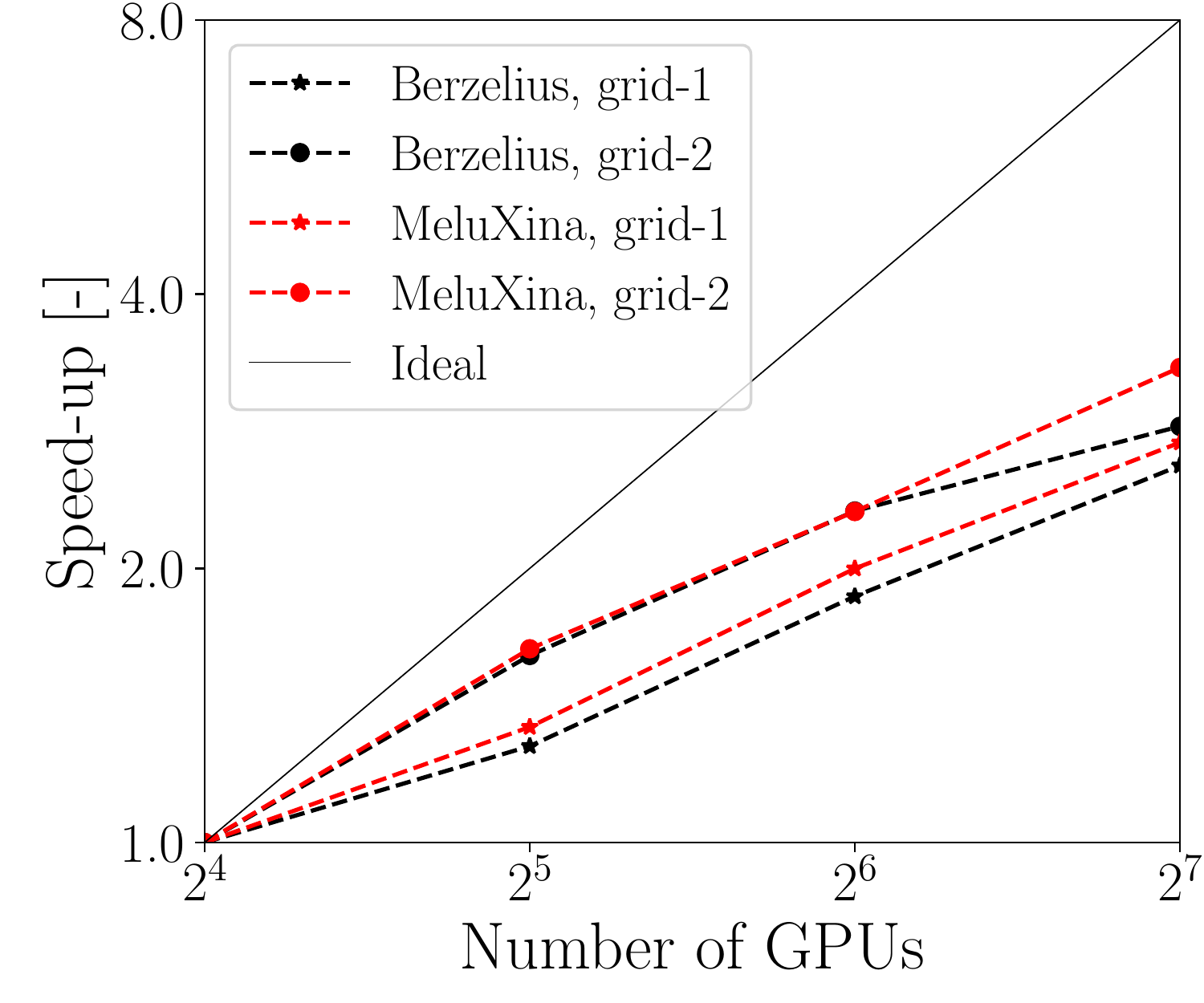}
	\caption{Strong scaling test performed on Berzelius (black-dashed lines) and MeluXina (red-dashed lines) clusters for two different grids: $1024\times 512\times 1024$ (grid-1) and $1024\times 1024\times 1024$ (grid-2). The black continuous line indicates the ideal behavior desired for the strong-scaling test.}
	\label{fig:strongS}
\end{figure}

We first discuss the weak-scaling tests for a Rayleigh-B\'enard problem with the same set-up as it will be  discussed in~\Cref{sec:trb}. For this test, we start with a "base" computational grid of $N_x\times N_y\times N_z=1024\times 512\times 256$ grid points on 2 GPUs. Then, while keeping fixed $N_x$ and $N_y$, we increase $N_z$ proportionally to the number of GPUs, resembling a procedure of spatial ensemble-average (i.e. more structures simulated to improve the convergence of the large-scale statistics). As discussed in~\Cref{sec:gpu_parallel}, we adopt a slab parallelization along the $z$ direction using the \code{-D\_DECOMP\_X} compilation option, which reduces to $2$ the number of \emph{all-to-all} operations. It is worth noticing that, although both HPC machines are equipped with A100-40GB NVIDIA cards, Berzelius has 8 GPUs/node while MeluXina has 4 GPUs/node. Moreover, the interconnection between GPUs is handled through NVLink, whereas node-to-node connection is performed through Infiniband (IB), known for having lower bandwidth and for operating on a different protocol. Hence, IB is less straightforward to handle as it requires a more careful configuration from a hardware and software perspective. This implies choosing the right MPI configurations and selecting compatible communication libraries, resulting in an efficiency that may vary significantly among different HPC centers. For these reasons, to prove the non IB-dependent scaling and maximize the GPU-to-GPU communication throughput, Berzelius was used to perform weak-scaling tests within a node. On the other hand, MeluXina was used for multiple-node tests to assess the IB-dependent scaling.\footnote{Berzelius uses a HGX 8-way platform, 8 A100 GPU connected all together, designed primarily for heavy AI workloads. MeluXina uses a HGX 4-way platform, 4 A100 GPU connected all together, which is better suited for scale-out HPC workloads. The majority of GPU-accelerated HPC clusters used primarily for simulation workloads in various scientific fields adopt the HGX 4-way configuration (including many European HPC systems funded by EuroHPC).}
\begin{figure}[h!]
    \centering
    \includegraphics[width=0.45\textwidth]{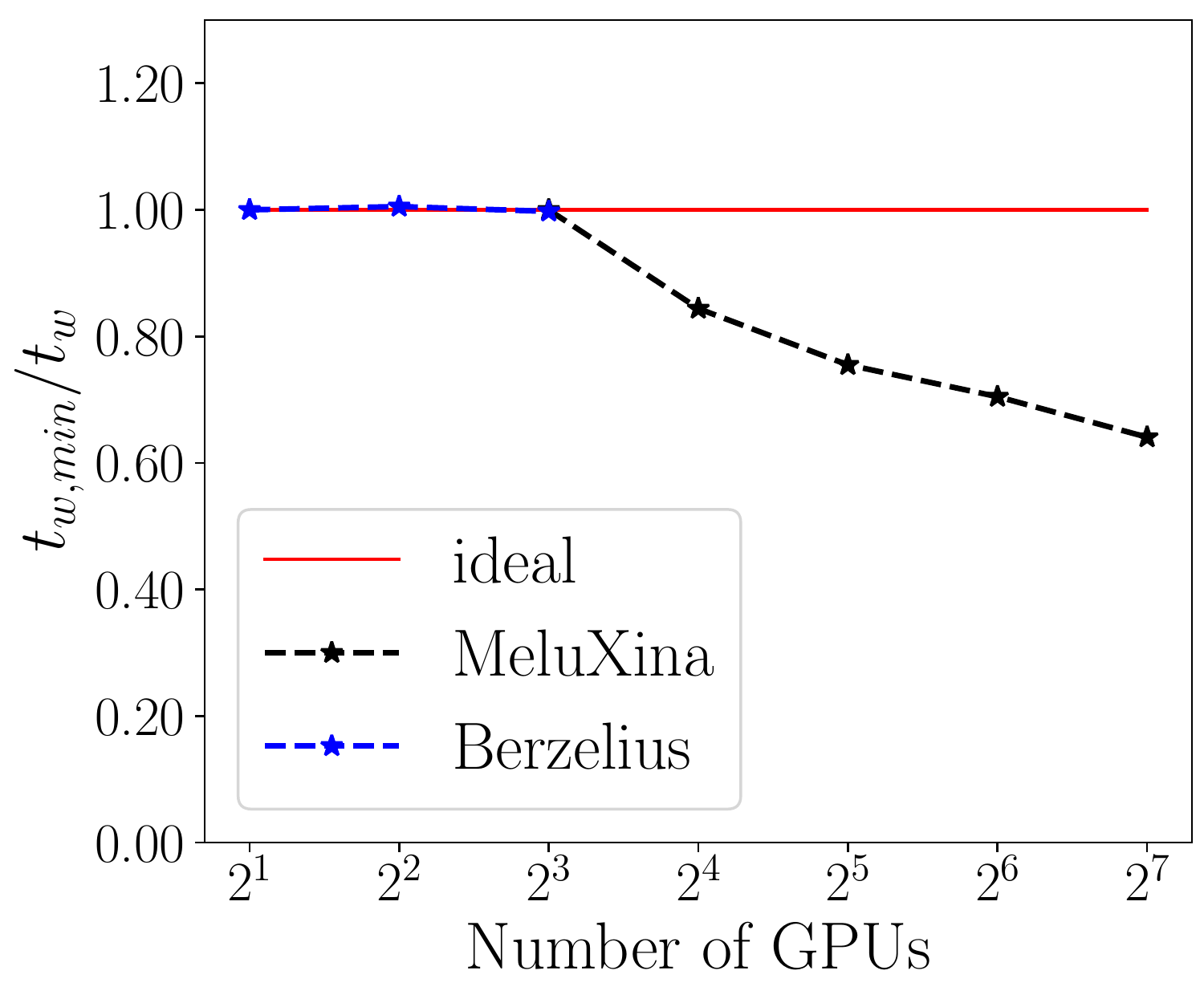}
    \includegraphics[width=0.45\textwidth]{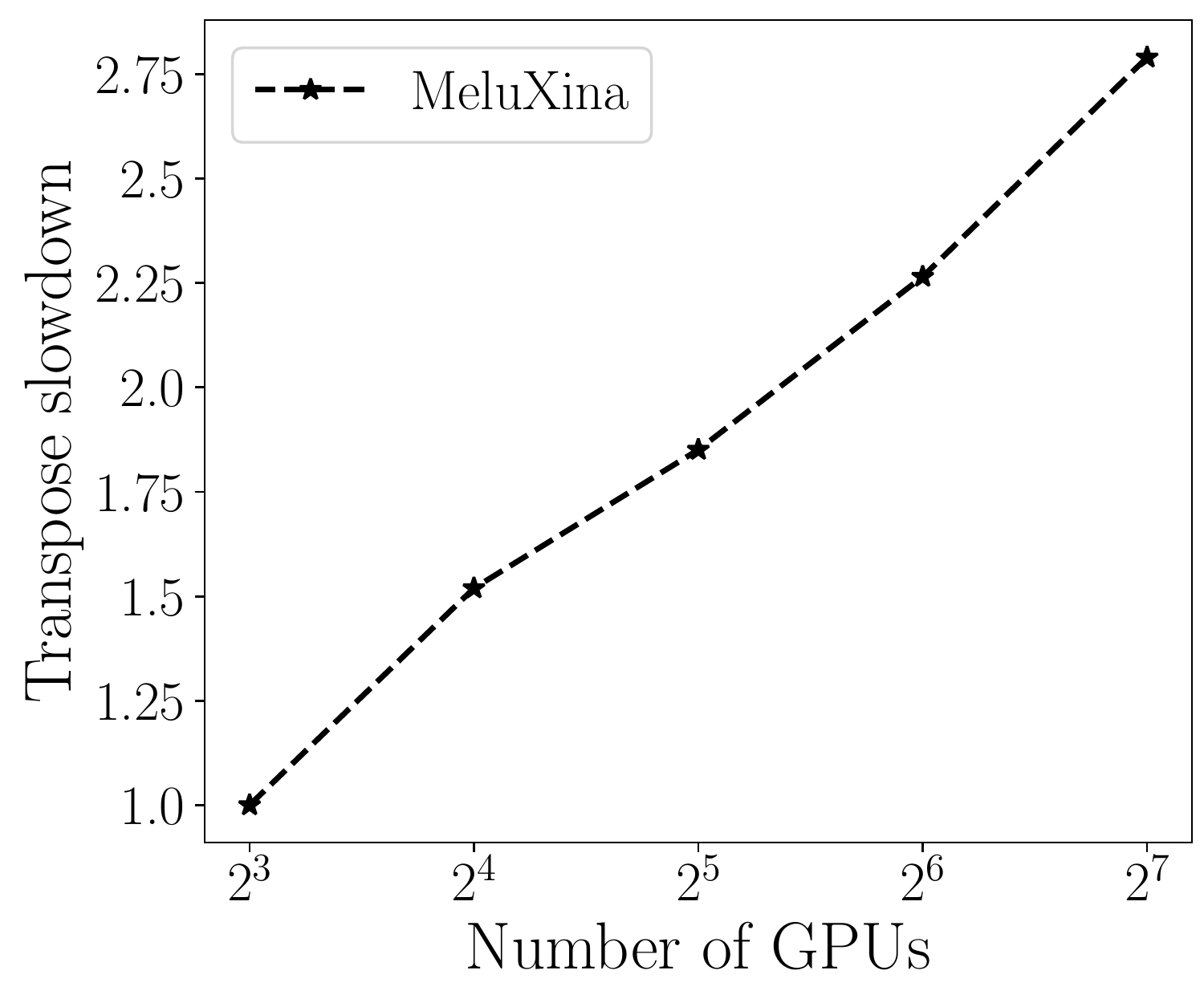}
    \put(-360,130){(\textit{a})}
    \put(-180,130){(\textit{b})}
    \caption{For the two-layer Rayleigh-B\'enard convection problem discussed in~\Cref{sec:trb}: a) code performance on MeluXina and Berzelius, b) slowdown due to transpose operation. For each set of data, we compute $t_{w,min}$ as the time per-timestep at the minimum number of GPUs tested over $t_{w}$, i.e. the time-per-timestep at the specific number of GPUs.}
    \label{fig:weaksc}%
\end{figure}
Figure~\ref{fig:weaksc}a) shows that weak-scaling is linear when bounded by NVLink communications (i.e. no IB communications), as clearly supported by tests on Berzelius. When IB communications are required (i.e. node-to-node data transfer) the code performances decrease. It is worth noticing that, while an increasing communication overhead is provided by node-to-node communication on IB network, additional slowdown is caused by the slab parallelization. By increasing the number of elements along $z$, more data need to be transferred during the \textit{x-to-z} transposes, further increasing the communication load. This is clearly shown in figure~\ref{fig:weaksc}b), where the slowdown is found to increase proportionally to the number of GPUs. \\
Results of the strong scaling tests are reported in figure~\ref{fig:strongS}. Here we use two different grids, \textit{i.e.} $1024\times 512\times 1024$ (grid-1) and $1024\times 1024\times 1024$ (grid-2) for the Rayleigh-B\'enard problem discussed in~\Cref{sec:trb}. Tests are performed on Meluxina and Berzelius as for the weak scaling. While keeping the problem sizes fixed, the number of GPUs is progressively increased up to a maximum of $128$, starting from $N_{GPU}=16$ which represents the minimum amount required to fit the two computational domains in the available GPU memory. Despite a speed-up is always achieved, the code shows a progressive loss in performance, i.e.\ a reduction of the benefits derived by increasing the number of GPUs. Note, however, that a larger number of grid point (e.g. grid-2) leads to lesser performance loss, as a higher GPU occupancy can be obtained. \\
The decrease in performance observed in figure~\ref{fig:strongS} is caused by two factors: the increase in communication among GPUs, and the reduction in local problem size, which does not leverage the full compute capacity of each GPU. While these effects are present in a strong scaling test, weak scaling allows us to isolate the effects of multi-GPU communication while keeping a higher GPU saturation. Thus, we argue that weak scaling represents a better tool to identify communication bottlenecks on multiple GPUs. Conversely, the strong scaling is more useful to estimate how much a fixed domain can be partitioned while keeping an efficient use of the computational resources. \\
%
%
Overall, the previous analysis suggests an important guideline for the user: in presence of unbalanced compute-vs-network architectures (e.g., node-to-node networking connection less efficient than the connection among GPUs within the same node), the optimal number of GPUs to be employed should be chosen as close as possible to the minimum amount required to fit the computational domain in the available GPU memory. Indeed, this is not always the case with older HPC architectures using previous generations of GPU hardware, where NVLink connections across GPUs inside a node were typically missing. For a fixed problem size, modern cards with high compute throughput will exhaust the required computation faster, leaving the remaining part of the computation as communication bound. In older GPU hardware, the acceleration is lower and communication becomes the dominant component affecting scalability for a larger number of GPUs. Hence, best practice dictates to use the least possible number of GPUs; on modern units with 80 GByte of HBM memory, if possible, it is therefore convenient to use one single 8-way GPU node (like a DGX A100) where all GPU are also connected via NVLink reducing communication overhead dramatically.
\subsection{CPU-GPU comparison}\label{subsec:cpu-gpu_comp}
In conclusion, we perform a comparison between the code performances on a CPU and a GPU architecture. It is worth mentioning that such comparison is notoriously not trivial. First, no exact and standard procedures to compare the two systems are established. Next, code performances may exhibit large variations among different architectures and using the same hybrid CPU-GPU node to perform tests on both may be misleading. CPU-only nodes and CPU-GPU nodes are intrinsically different in terms of network configuration and GPU/CPU interconnection, hence an unbiased test may not be performed directly on hybrid architectures (as CPU-GPU cluster would hardly be used to perform CPU-only jobs). Therefore, the following analysis has to be taken as a first-approximation estimate. \\
\begin{figure}[h!]
	\centering
	\includegraphics[width=0.49\textwidth]{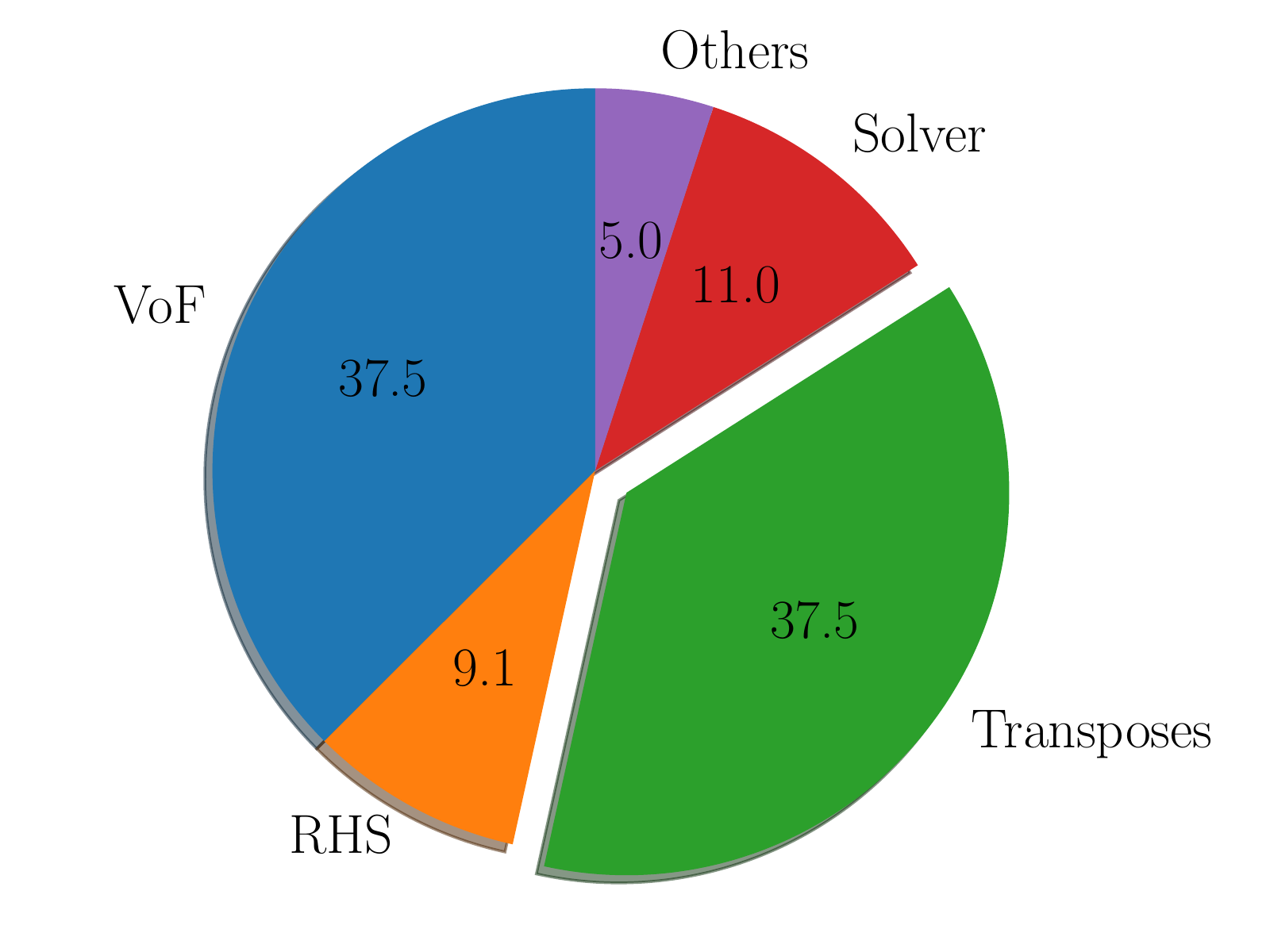}
	\put(-190,140){(\textit{a})}
	\includegraphics[width=0.49\textwidth]{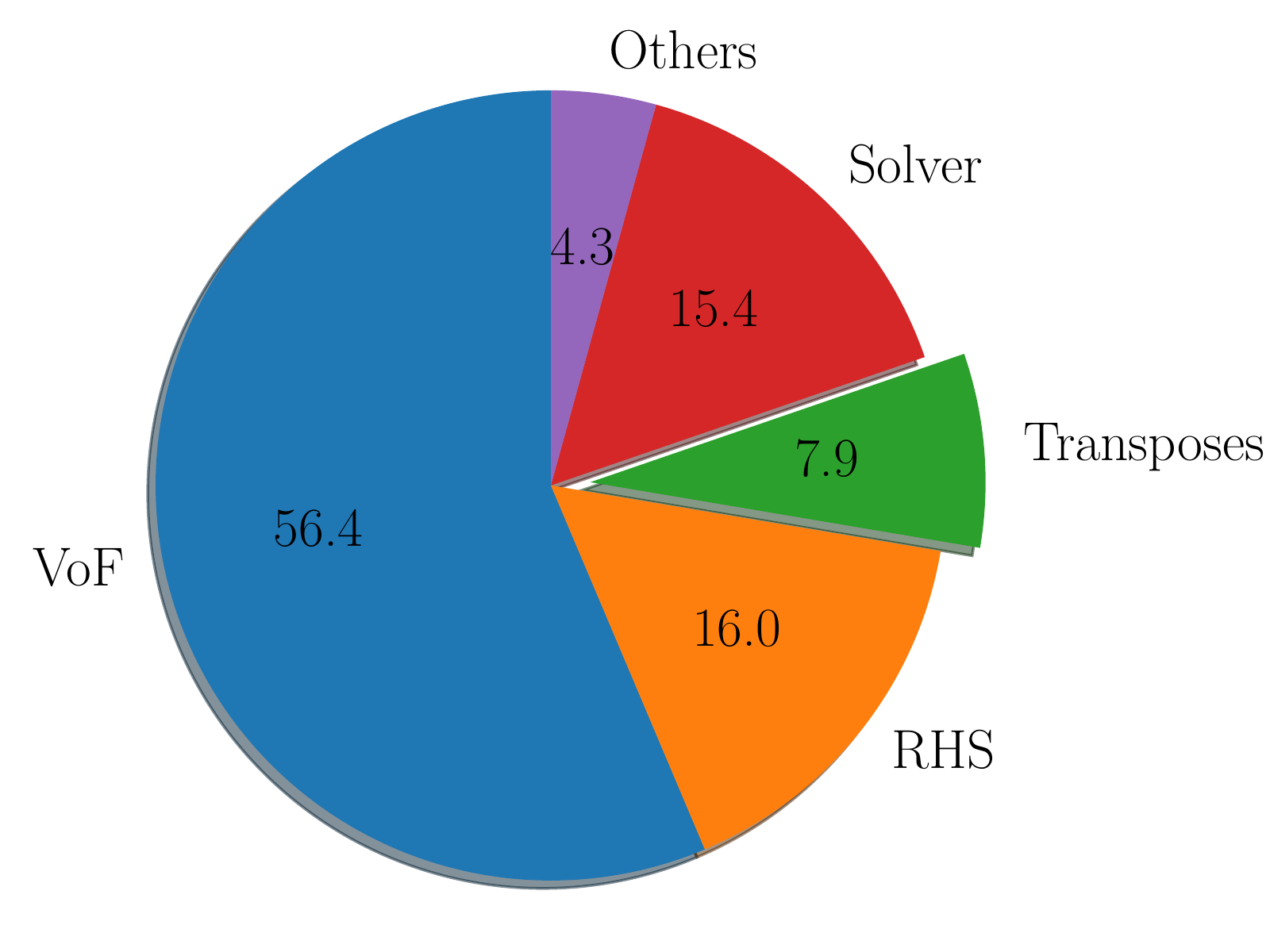}
	\put(-190,140){(\textit{b})}
	\caption{Comparison of code-section load percentage on the total simulation time for GPUs (panel \textit{a}) and CPUs (panel \textit{b}). The different "slices" represent different code sections: 1) VoF (i.e. interface reconstruction and advection, update of the thermophysical properties), 2) RHS (i.e. discretization of the governing equations), 3) Transposes (i.e. transpose operation in the solver), 4) Solver (i.e.\ only Gaussian elimination) and others (i.e. correction step, divergence/time-step checks, output and post-processing routines).}
	\label{fig:cpu-gpu-comp}
\end{figure}%
Here, we repeat the weak-scaling simulation with $n_{GPU}=8$ GPUs on Berzelius on $n_{CPU}=512$ CPUs on Tetralith, in both cases employing a slab parallelization along $z$. The test shows that for GPUs the average wall-clock time per-timestep is $t_{8,GPU}=0.191$ $s$, while for CPUs $t_{512,CPU}=1.075$ $s$. This results in an equivalent number of GPUs $n_{eq}=(t_{512,CPU}n_{CPU})/(t_{8,GPU}n_{GPU})\approx$ $359$. \\
Finally, a comparison in terms of computing-load percentage for each code section is displayed in figure~\ref{fig:cpu-gpu-comp}. As previously anticipated, transposes during GPU simulations (panel \textit{a}) represents more than half of the computing load. The remaining parts, mainly composed of stencil operations enclosed in \code{for} loops, largely benefit from GPU-offload, while in CPUs (panel \textit{b}) these account for more than 70 \% of the total wall-clock time per time-step.

\section{Applications}\label{sec:app_2}
\subsection{Emulsions in HIT}\label{sec:hit} 
\begin{figure}[t!]
	\centering
	\includegraphics[width=0.49\textwidth]{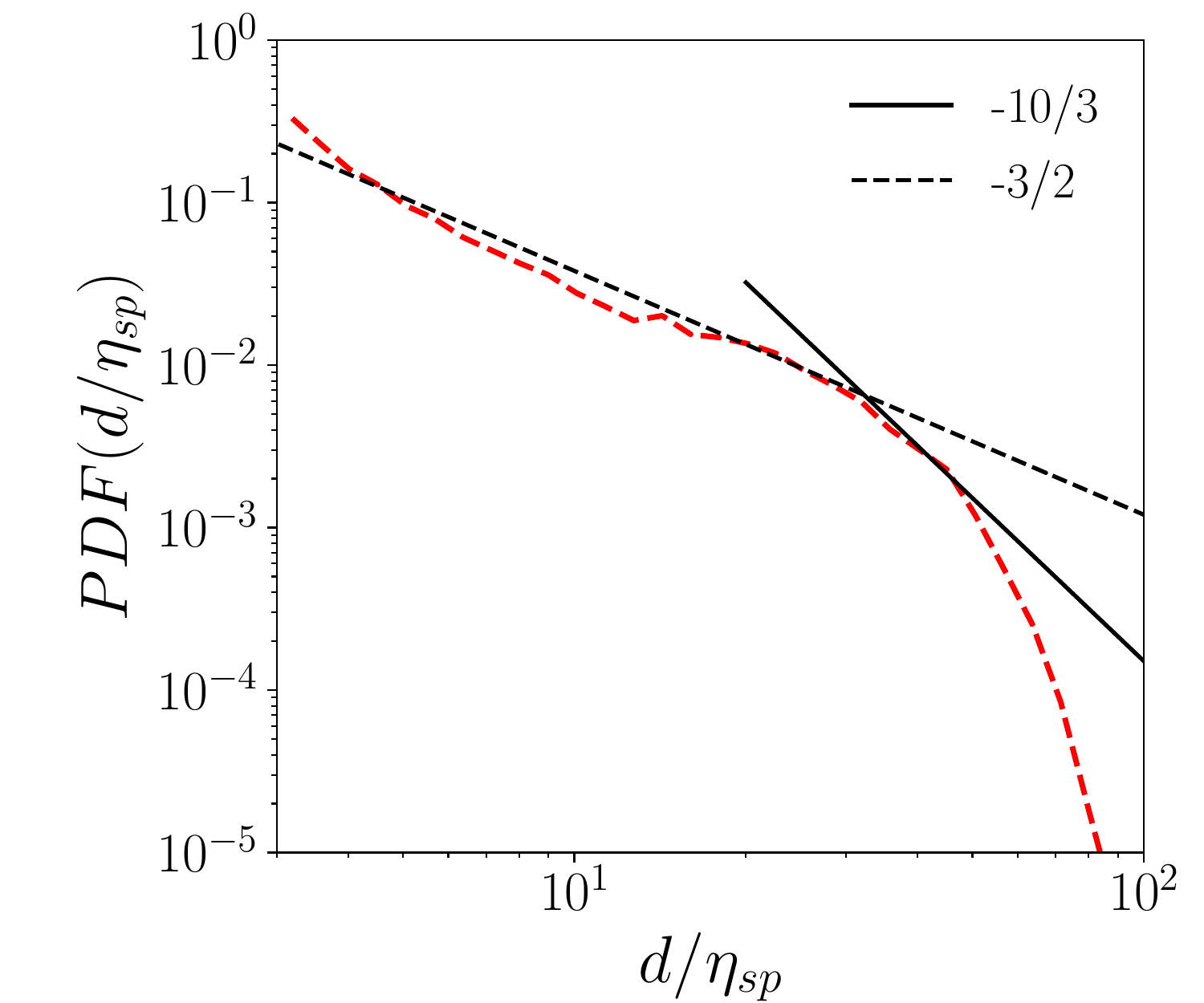}
	\put(-190,140){(\textit{a})}
	\includegraphics[width=0.49\textwidth]{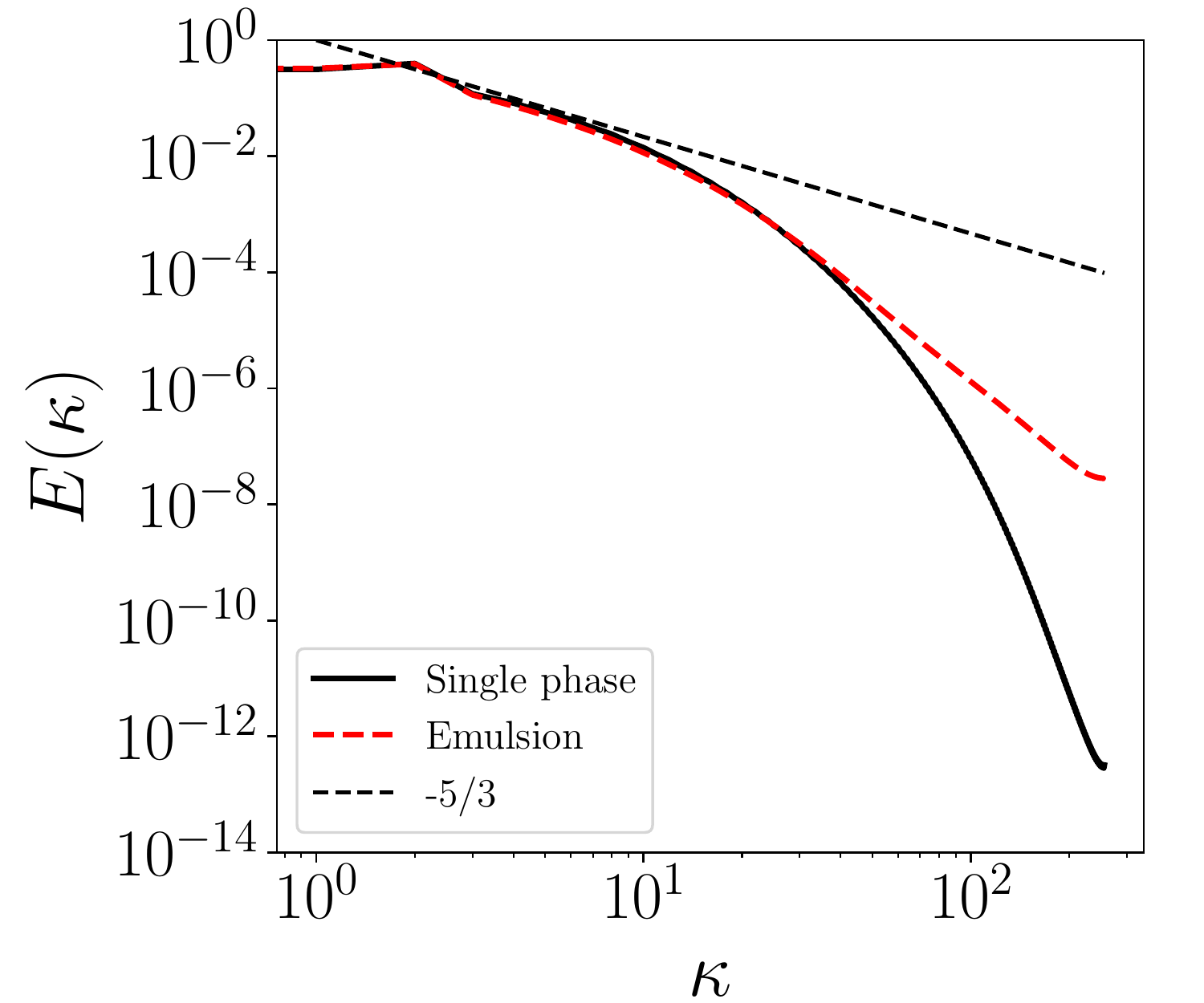}
	\put(-190,140){(\textit{b})}
	\caption{Results at statistical-stationary state for turbulent emulsions simulation. Panel (\textit{a}) shows the droplet size distribution, where the droplet diameter $d$ is normalized by the single-phase Kolmogorov scale $\eta_{sp}$. Also in black, the $-10/3$ and $-3/2$ power-laws are shown (see \cite{Garrett2000,Deane2002}). Panel (\textit{b}) shows the one-dimensional turbulent kinetic energy spectra, comparing the single-phase and multiphase (i.e. emulsion) cases. Here, the $-5/3$ law for the inertial range is shown, which applies for almost a decade.}
	\label{fig:hit2}%
\end{figure}

In many multiphase flows, the dispersed phase interacts with the surrounding turbulence induced through large-scale stirring mechanisms. While turbulence introduces a vast range of scales, usually spanning over several order of magnitudes, the presence of an interface introduces further complexity, offering alternative paths for energy transmission through scales and generating poly-dispersed droplet/bubble distributions. Resolving the interplay between all these mechanisms leads to an extremely complex scenario to simulate numerically. Even in simplified conditions, represented by Homogeneous and Isotropic Turbulence (HIT), the number of grid points could rapidly exceed $N \geq 1024^3$, quite challenging for multiphase flows. Furthermore, variations of density and viscosity, and variations of the surface tension coefficient may introduce smaller scales as lower viscosity in the dispersed phase may accelerate vortices.
\begin{table}[h]
	\centering
	\begin{tabular}{cccccc}
		\hline
		 $Re_\lambda$ & $We_\mathcal{L}$& $\mathcal{V}$ & $\mathcal{L}$ & $\rho_1/\rho_2$ & $\mu_1/\mu_2$ \\
		\hline
		137    &    42.6           &       0.06       &    $\pi$	    &   1         &    1 \\	
		\hline
	\end{tabular}
	\caption{Physical dimensionless parameters of the present configuration: $Re_\lambda=u_{rms}\lambda/\nu_2$ and $We_\mathcal{L}=\rho_1u_{rms}^2\mathcal{L}/\sigma$ are the corresponding Taylor microscale Reynolds number and the large scale Weber number, $\mathcal{L}$ is the large-scale at which turbulence is forced, $\mathcal{V}$ is the volume fraction (ratio between the volume occupied by the disperse phase and the total volume). The simulation is performed at matching density and viscosity (i.e., $\rho_1/\rho_2=\mu_1/\mu_2=1$).}
	\label{table:hit}
\end{table}
\begin{figure}[h!]
	\centering
	\includegraphics[width=0.6\textwidth]{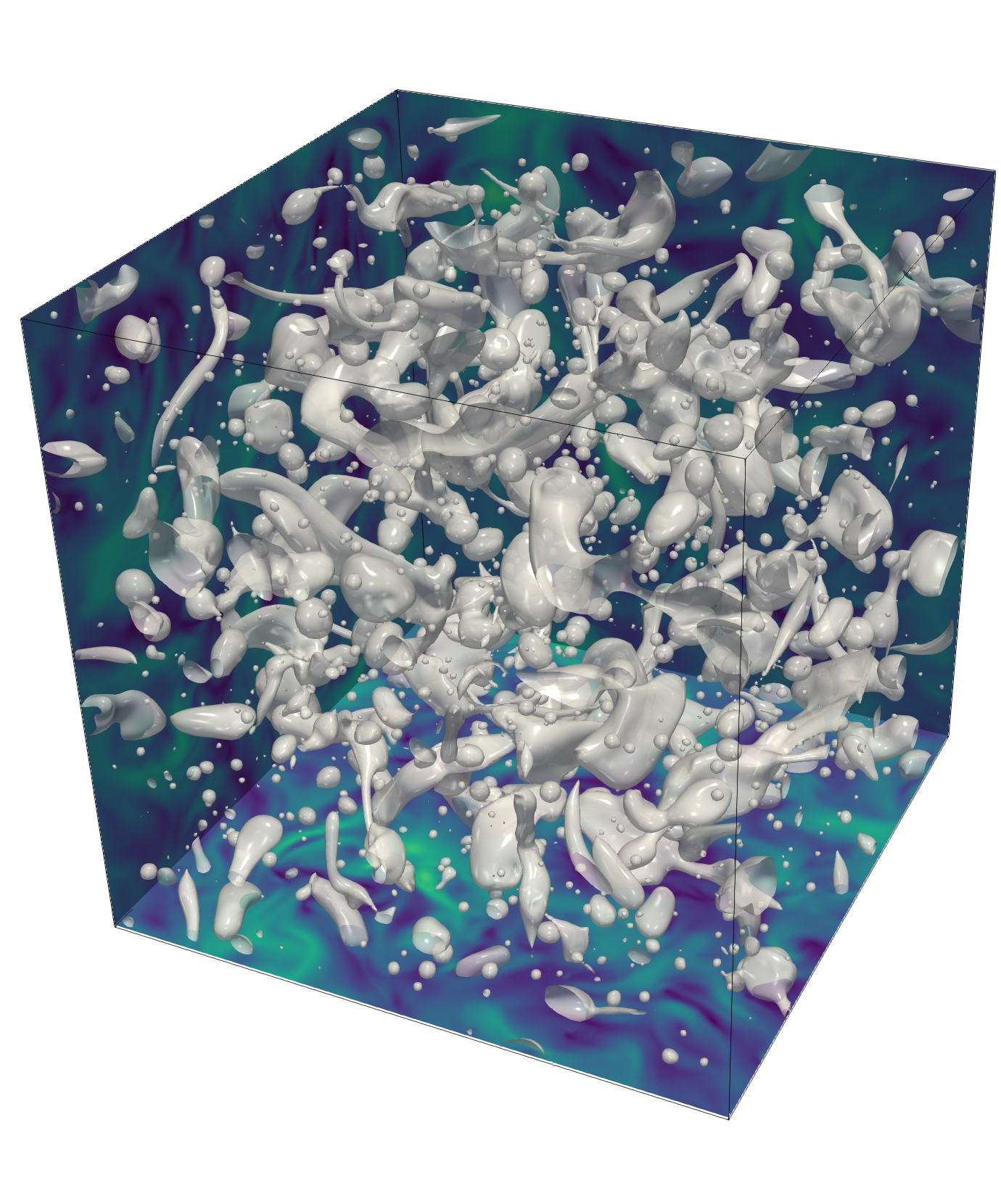}
	\caption{Render for the iso-contour of the volume fraction value $\phi=0.5$. Lateral planes show the modulus of the vorticity field.}
	\label{fig:hit1}%
\end{figure}
Fully developed turbulence is usually reached for Taylor-Reynolds number $Re_\lambda=u_{rms}\lambda/\nu\gtrsim 200$, where $\lambda$ is the Taylor scale and $u_{rms}$ is the root-mean-square fluctuation velocity. In multiphase flow simulations, these intensities are rarely reached as typical values are $Re_\lambda \lesssim 100$. In this example, we present a simulation of a turbulent emulsion at $Re_\lambda\approx 137$ performed on a grid of $512^3$ on a cube of side length $2\pi$ and with the main physical parameters reported in table~\ref{table:hit}. Turbulence is sustained at large scale using the Arnold-Beltrami-Childress forcing (see \cite{Podvigina1994,Mininni2006}) throughout the whole simulation. \\
Figure~\ref{fig:hit1} shows a render at statistically-stationary state. Turbulence is first simulated in single phase; when statistically-stationary conditions are reached, the dispersed phase is introduced through a random distribution of droplets and let to develop until convergence is reached. This is monitored in terms of droplet-size-distribution and one-dimensional spectra, see figure~\ref{fig:hit2}.
Despite the simplicity of this configuration, HIT offers a relevant framework to study multiphase turbulence in complex configurations. To reach properly convergent statistics, several large-eddies turnovers are required, corresponding to $\sim 10^7$ time-steps. Hence, GPU acceleration will be invaluable to reach fully-developed turbulent conditions in future studies.

\subsection{Two-layer Rayleigh-B\'enard convection}\label{sec:trb} 

Rayleigh--B\'enard convection is the flow developed inside a fluid layer that is heated from below and cooled from above. It is driven by the density differences that arise due to the temperature variation inside the fluid. Even though it is a seemingly simple configuration, it encapsulates rich physics that are encountered in a range of engineering applications and physical phenomena.  Beyond the classical setting, the study of the two-layer Rayleigh--B\'enard variant is crucial from both a fundamental and an applied standpoint. First, regardless of the application, there is always some dissolved gas in every liquid. Therefore, it is inevitable that a gaseous phase will be formed in any realistic natural convection flow. Second, physical phenomena such as the convection in the earth’s mantle~\cite{busse1981aspect} or engineering applications such as the heat transfer inside magnetic confinement systems in fusion reactors~\cite{wilczynski2019stability} are more accurately modelled as two-layer convection, where the two fluid layers are dynamically coupled. 

\begin{figure}[t!]
    \centering
    \includegraphics[width=0.8\textwidth]{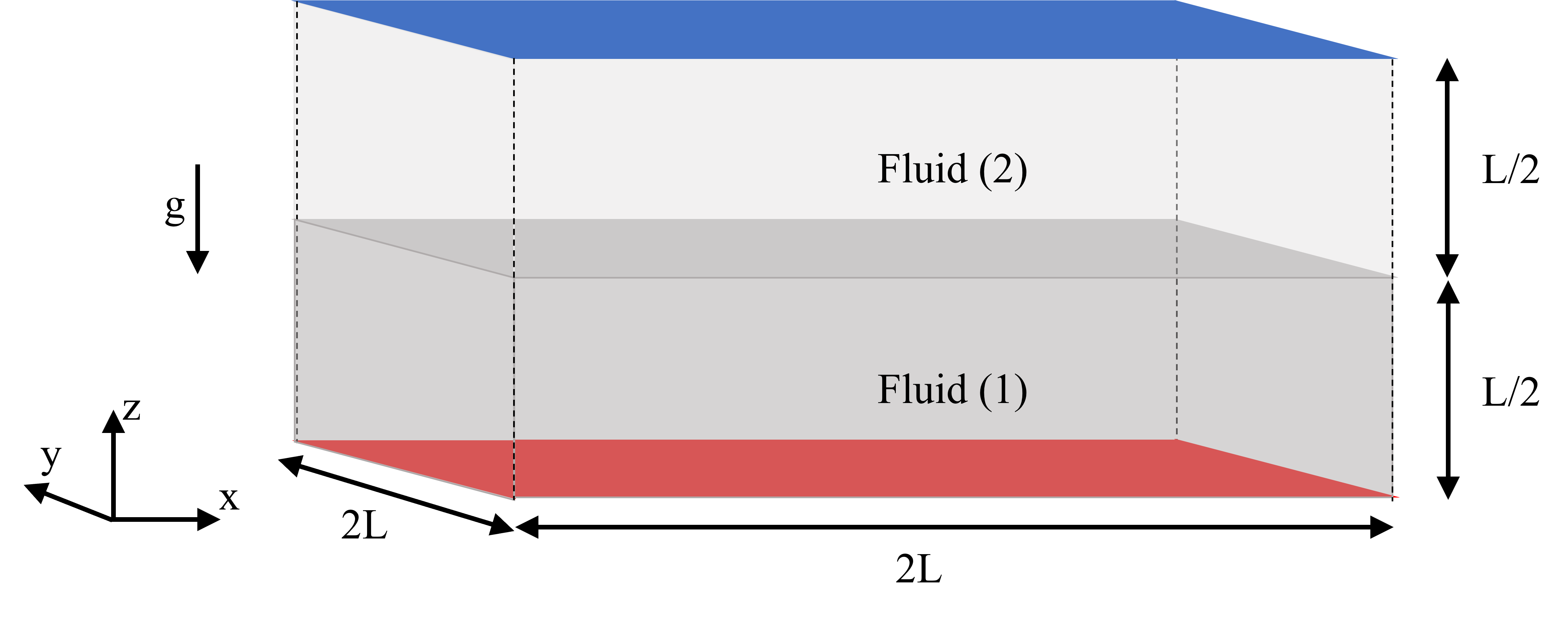}
    \caption{Schematic representation of the geometry used for the two-layer Rayleigh--B\'enard convection. The bottom heated surface is depicted in red and the top cooled surface in blue.}
	\label{fig:two_layer_tmp_geometry}
\end{figure}

Figure~\ref{fig:two_layer_tmp_geometry} shows a schematic representation of the domain used for the numerical simulations, as  used previously in~\cite{liu2021two}, in two dimensions. The bottom and top walls are modeled as solid isothermal surfaces at constant temperatures of 328~K and 318~K respectively. The x- and y-directions are considered periodic, and the aspect ratio between the horizontal and vertical dimensions of the cavity is $\Gamma=L_x/L_z=L_y/L_z=2$. The dimensionless parameters adopted are shown in Table~\ref{table:parameters}. The property ratios between the two fluids are considered equal to 1 (kinematic viscosity $\nu$, thermal diffusivity $\alpha$, specific heat $c_p$ and thermal expansion coefficient $\beta$), except the density ratio $\lambda_{\rho}=\rho_2/\rho_1=0.1$. This mismatch in densities is the reason behind the arrangement of the fluids in a two-layer configuration. Preliminary simulations revealed that a grid of $1024\times1024\times512$ cells and a CFL number of 0.50 were adequate for obtaining grid- and time step-independent solutions.

\begin{table}[t]
\centering
\begin{tabular}{lccccc}
\hline 
 $\lambda_{\rho}=\frac{\rho_2}{\rho_1}$ &
 Pr=$\frac{\nu_1}{\alpha_1}$ &
 Ra=$\frac{|\mathbf{g}| \beta_1 \Delta T L^3}{\nu_1 \alpha_1}$ & 
 We= $\frac{\rho_1 |\mathbf{g}| \beta_1 \Delta T L^2}{\sigma}$ &
 Fr=$\sqrt{\beta_1 \Delta T}$  \\ 
\hline
0.1 & 1 & $10^6,10^7,10^8$ & $100$ & $1$ \\
\hline
\end{tabular}
\caption{Dimensionless parameters adopted for the study of two-layer Rayleigh--B\'enard convection. All other property ratios are equal to 1.}
\label{table:parameters}
\end{table}

\begin{figure}[t!]
  \centering
  \includegraphics[width=0.8\textwidth]{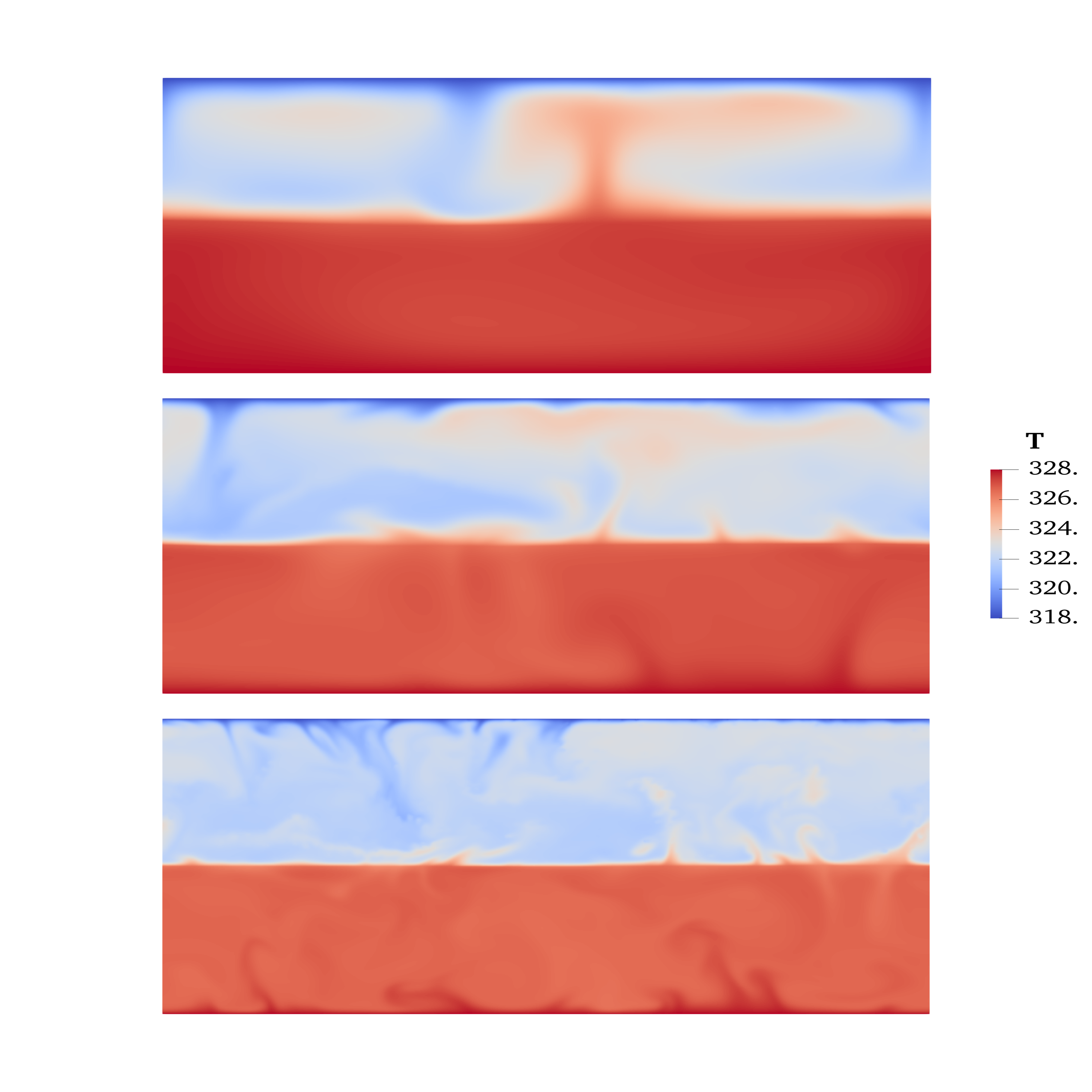}\\
  \begin{picture}(0,0)(0,0)
    \put(-130,380){(\textit{a})}
    \put(-130,260){(\textit{b})}
    \put(-130,140){(\textit{c})}
  \end{picture} 
  \caption{Instantaneous temperature fields in the x--y plane for the two-layer Rayleigh--B\'enard convection. (\textit{a}) $Ra=10^6$, (\textit{b}) $Ra=10^7$, (\textit{c}) $Ra=10^8$. }
  \label{fig:two_layer_tmp_2D}
\end{figure}

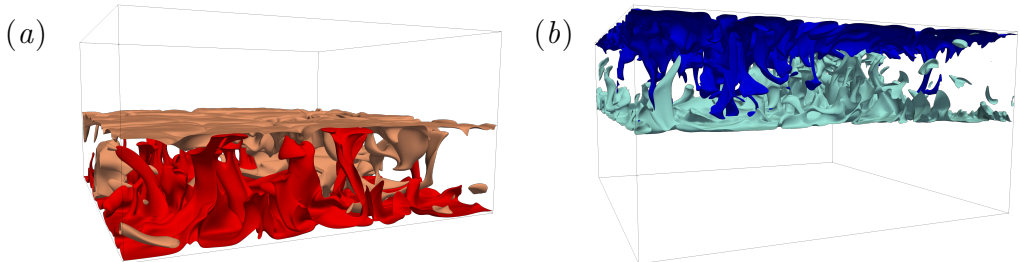
\begin{figure}[t!]
    \centering
    \includegraphics[width=0.49\textwidth]{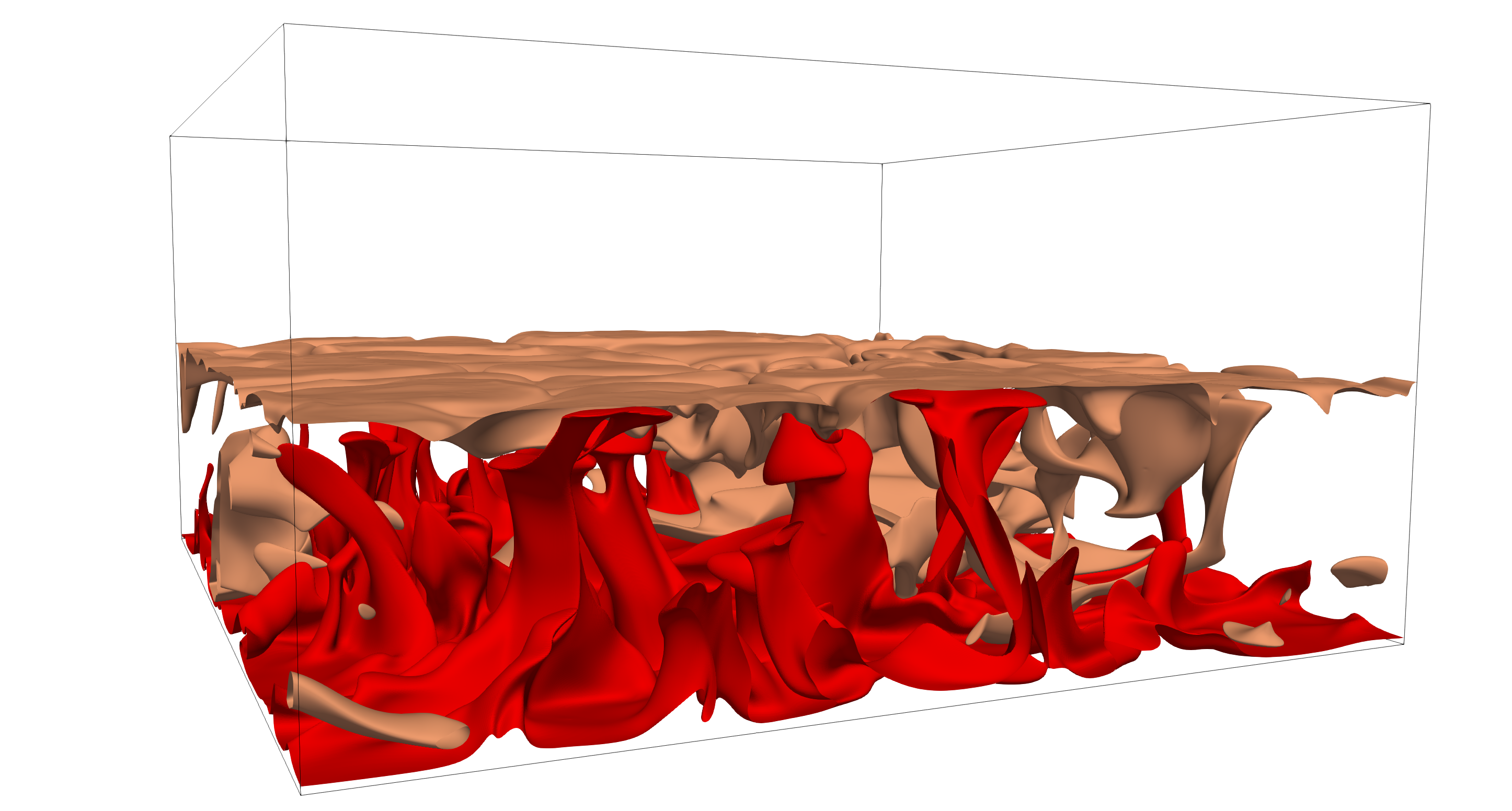}
    \includegraphics[width=0.49\textwidth]{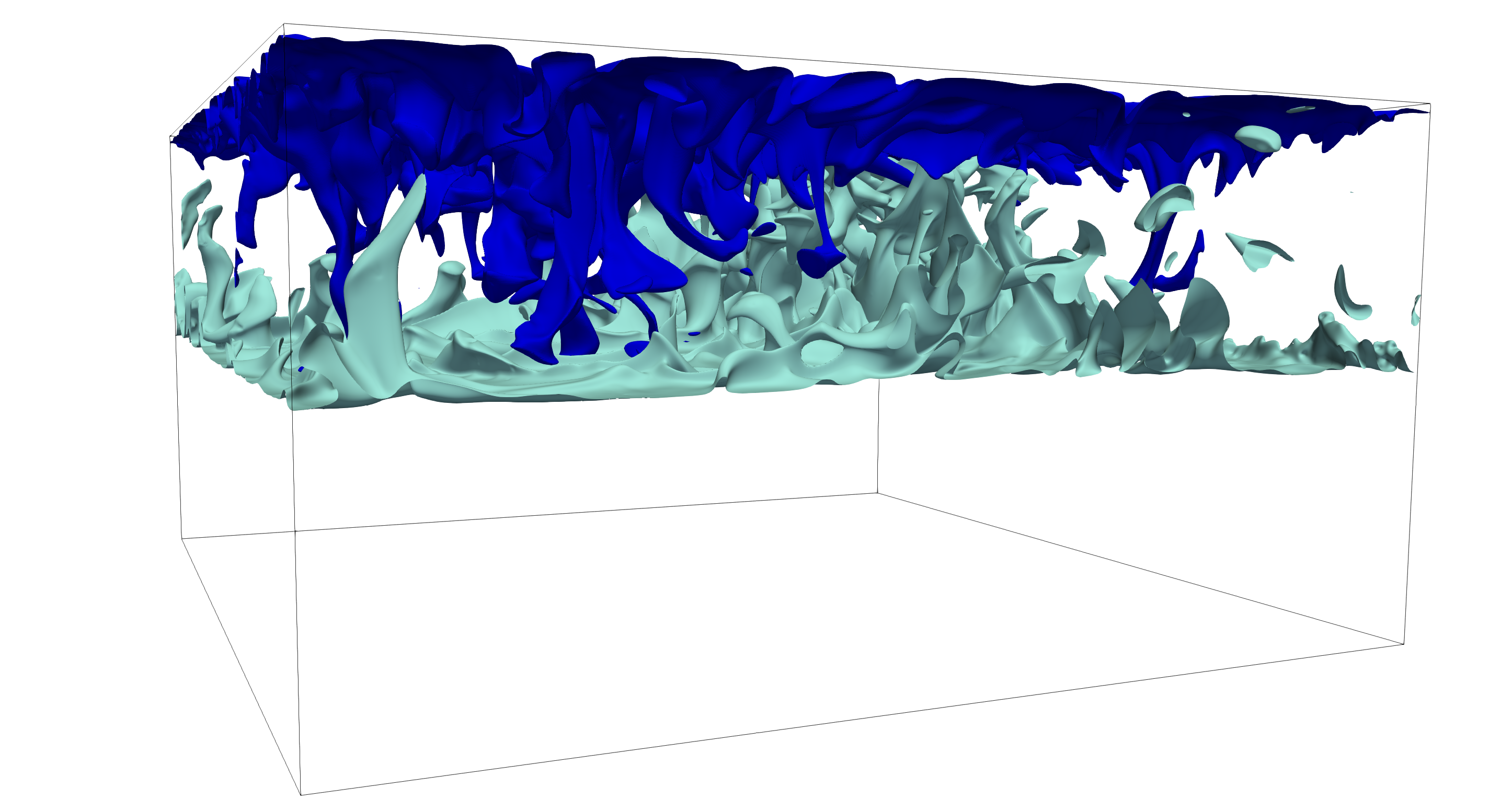} \\
    \begin{picture}(0,0)(0,0)
        \put(-200,100){(\textit{a})}
        \put(0,100){(\textit{b})}
    \end{picture} 
    \caption{Instantaneous temperature contours for the two-layer Rayleigh--B\'enard convection at $Ra=10^8$. (\textit{a}) Bottom half of the cavity with red plumes at 327~K and orange plumes at 326.5 K. (\textit{b}) Top half of the cavity with cyan plumes at 323~K and blue plumes at 321~K. }
	\label{fig:two_layer_tmp_3D}
\end{figure}

Figure~\ref{fig:two_layer_tmp_2D} shows instantaneous temperature fields in the x--y plane for the three Rayleigh numbers considered. With increasing Rayleigh number, the thermal structures in the cavity become finer, indicating increased turbulent activity. This increased thermal agitation is not enough to induce any significant interface movement, even at the highest Rayleigh number considered. Furthermore, in all three cases, the temperature drop at the bottom fluid layer is much smaller than the top layer. This is explained by the fact that the two layers have the same thermal diffusivity and specific heat but different densities, leading to a top layer with a smaller thermal conductivity ($k=\alpha \rho c_p$) compared to the bottom layer. Therefore, the top wall conducts heat less effectively than the bottom wall, which explains the larger temperature gradients at the top layer.
Focusing on the highest Rayleigh number case, figure~\ref{fig:two_layer_tmp_3D} shows instantaneous temperature contours for $Ra=10^8$. As in classical Rayleigh--B\'enard convection, hot and cold plumes are ejected from the bottom and top walls. In the presence of a single fluid layer, these plumes would typically get organized in large scale circulation structures, extending from the bottom to the top wall. Here, the existence of two fluid layers changes the classical picture; the interface acts as a barrier, confining the thermal plumes in each fluid layer. The interface also acts as a thermal conductor, promoting the exchange of heat between the two layers. More specifically, the hot plumes ascending from the bottom wall are cooled off when reaching the interface, forming colder plumes that travel downwards. The opposite is true at the top layer, where the descending cold plumes are heated by the interface and hotter plumes emerge from the interface traveling upwards. Figure~\ref{fig:two_layer_tmp_3D} clearly illustrates this behavior, revealing the existence of regions dominated by ascending plumes and regions dominated by descending plumes, hinting towards the organization of the flow in three-dimensional large scale circulation structures in each layer.

\section{Conclusions and further developments}\label{sec:concl} 
We present the code FluTAS, a numerical framework tailored for direct numerical simulations of multiphase flows, with the option of heat transfer, able to run efficiently on CPU-based standard architectures and on GPU-based accelerated machines. The open source version, released under MIT license, includes a pressure-correction algorithm for two-phase flows extended with an algebraic Volume-of-Fluid method (MTHINC) for capturing the interface dynamics. \\
We provide here a description of the employed numerical algorithm, with details on the solution of the governing equations and of the advection of the interface. After presenting different validation benchmarks both in single and multiphase configurations, we discuss the code performance focusing on two aspects: i) its current limitation when the communications among GPUs in different nodes are considerable less efficient than the communication among GPUs within the same node, ii) its advantages compared to CPUs in terms of "time-to-solution". Finally, we report results from two configurations of fundamental interests in multiphase turbulence: emulsions in homogeneous isotropic turbulence and the two-layer Rayleigh-B\'enard convection. \\
In the future, we aim to improve the code maintainability and portability (both on CPU and GPU) and to release additional modules under development, e.g.\ weak compressibility and phase change~\cite{dalla2021interface,scapin2022finite}. Further efforts will be devoted to enhance the code performance on multiple GPUs nodes, reducing the current communication bottlenecks. To this end, a promising strategy is the one proposed in~\cite{ha2021multi}, i.e., implement the solution of the tridiagonal system for the third direction on a distributed memory. The main advantage of this approach is the elimination of \emph{all-to-all} operations in the Poisson solver. This improvement combined with future enhancements in the software frameworks for collective data communications will allow tackling several multiphase problems while keeping an efficient use of the computational resources.

\section*{Acknowledgements}
We would also like to thank the staff at CINECA, notably Massimiliano Guarrasi and Fabio Pitarri for their help in the development of the GPU functionalities at an early stage of the work. We thank Francesco De Vita for the help with the first implementation of the VoF MTHINC method.

\section*{Funding}
M.C-E., N.S., A.D. and L.B. acknowledge the support from the Swedish Research Council via the multidisciplinary research environment INTERFACE, Hybrid multiscale modelling of transport phenomena for energy efficient processes and the Grant No.~2016-06119. M.E.R. was supported by the JSPS KAKENHI Grant Number JP20K22402. P.C. was supported by the University of Iceland Recruitment Fund grant No.~1515-151341, \emph{TURBBLY}. The computer time was provided by SNIC (Swedish National Infrastructure for Computing), by the National Infrastructure for High Performance Computing and Data Storage in Norway (project No.~NN9561K) and by CINECA (Marconi100) in Italy. Scaling tests on multiple GPUs has been performed on Berzelius (under the project No.~Berzelius-2022-29 and operated by SNIC) and MeluXina (under the project number No.~EHPC-BEN-2022B01-027 with the EuroHPC Benchmark Access and managed by LXP). M.E.R. acknowledges the computer time provided by the Scientific Computing section of Research Support Division at OIST.

\appendix

\section{Implementation details of MTHINC method}\label{sec:details_mthinc} 
\subsection{Calculation of the reconstructing polynomial}
In the current work, we consider a polynomial up to second order and, therefore, $\mathcal{T}$ can be expressed using the following general quadratic form:
\begin{equation}
    \mathcal{T}(\tilde{\mathbf{x}}) = \mathbf{c}\cdot\left[\mathcal{Q}_{\mathcal{T},1}\cdot\left(\tilde{\mathbf{x}}\cdot\tilde{\mathbf{x}}^T\right)^T\right]+\mathbf{c}\cdot\mathbf{c}^R\cdot\mathcal{Q}_{\mathcal{T},2}\cdot\left(\tilde{\mathbf{x}}\cdot\tilde{\mathbf{x}}^R\right)^T+\mathcal{L}_{\mathcal{T}}\cdot\tilde{\mathbf{x}}^T\mathrm{.}
  \label{eqn:qud_fun}
\end{equation}
where $\tilde{\mathbf{x}}^R=(\tilde{y},\tilde{z},\tilde{x})$ while the components of the flag vectors (equal to $0$ or $1$) $\mathbf{c}$ and $\mathbf{c}^R$ will be discussed later in~\ref{subsec:d_th}. Accordingly, to determine $\mathcal{T}$, one needs just to compute the components of the vectors $\mathcal{Q}_{\mathcal{T},i=1,2}$ and $\mathcal{L}_{\mathcal{T}}$. This can be done by first imposing that the first-order and second-order gradient of $\mathcal{T}$, evaluated for $\tilde{\mathbf{x}}=\tilde{\mathbf{x}}_c$, are equal to the normal vector $\mathbf{n}$ and the curvature tensor $\mathbf{K}$:
\begin{equation}
  \begin{cases}
    \left.\nabla\mathcal{T}\right|_{\tilde{\mathbf{x}}=\tilde{\mathbf{x}}_c} &= \mathbf{n}\mathrm{,} \\
    \left.\nabla\left(\nabla\mathcal{T}\right)\right|_{\tilde{\mathbf{x}}=\tilde{\mathbf{x}}_c} &= \mathbf{K}\mathrm{.}
  \end{cases}
    \label{eqn:constrain}
\end{equation}
Expanding equation~\ref{eqn:qud_fun} and the equations~\ref{eqn:constrain} lead after some manipulation to a unique expression for the components of the vectors $\mathcal{Q}_{\mathcal{T},i=1,2}$ and $\mathcal{L}_{\mathcal{T}}$:
\begin{equation}
  \mathcal{Q}_{\mathcal{T},1} = \left[a_{Q,1}^x,a_{Q,1}^y,a_{Q,1}^z\right] = \dfrac{1}{2}\left[K^{xx},K^{yy},K^{zz}\right]\mathrm{,}
  \label{eqn:q1_comp}
\end{equation}
\begin{equation}
  \mathcal{Q}_{\mathcal{T},2} = \left[a_{Q,2}^x,a_{Q,2}^y,a_{Q,2}^z\right] = \left[K^{xy},K^{yz},K^{xz}\right]\mathrm{,}
  \label{eqn:q2_comp}
\end{equation}
\begin{align}
  \mathcal{L}_{\mathcal{T}} = \left[a_L^x,a_L^y,a_L^z\right] = \,\, & \hspace{0.000 cm} \left[ n^x-\dfrac{c^x}{2}\left(K^{xx}+c^yK^{xy}+c^zK^{xz}\right)\mathrm{,}\right. \nonumber \\
                                                               \,\, & \hspace{0.195 cm} \left. n^y-\dfrac{c^y}{2}\left(K^{yy}+c^xK^{xy}+c^zK^{xz}\right)\mathrm{,}\right. \nonumber \\
                                                               \,\, & \hspace{0.195 cm} \left. n^z-\dfrac{c^z}{2}\left(K^{zz}+c^xK^{xz}+c^yK^{yz}\right)\right]\mathrm{.}
  \label{eqn:l_comp}
\end{align}
Note that to recover a linear reconstruction, the curvature tensor is set identically zero and thus also the components of the vector $\mathcal{Q}_{\mathcal{T},i=1,2}$ in equations~\eqref{eqn:q1_comp} and~\eqref{eqn:q2_comp}. Accordingly, from equation~\eqref{eqn:l_comp}, $\mathcal{L}_{\mathcal{T}}=[n^x,n^y,n^z]$ and equation~\eqref{eqn:qud_fun} reduces to $\mathcal{T}(\tilde{\mathbf{x}})=\mathcal{L}_{\mathcal{T}}\cdot\tilde{\mathbf{x}}^T$ which represents the equation of a plane in the three-dimensional Cartesian space.

\subsubsection{Normal vector and curvature calculation}\label{subsec:normal_kappa}
Given the smooth nature of the color function, the normal vector $\mathbf{n}$ and the curvature tensor $\mathbf{K}$ can be computed directly from the corresponding geometrical definitions:
\begin{equation}
  \mathbf{n} = \dfrac{\nabla\phi}{|\nabla\phi|}\mathrm{,}
  \label{eqn:normal}
\end{equation} 
\begin{equation}
  \mathbf{K} = -\nabla\mathbf{n}\mathrm{,}
  \label{eqn:kappa_t}
\end{equation} 
where $\mathbf{m}=\nabla\phi=(m^x,m^y,m^z)$. Following the Youngs' method~\cite{youngs1982time,youngs1984interface}, the three components $m^x$, $m^y$ and $m^z$ (the partial derivative of $\phi$ in each direction) are computed by first evaluating the derivatives at the cell corners and then each corner-value is averaged to find $\partial\phi/\partial x$, $\partial\phi/\partial y$ and $\partial\phi/\partial z$. Once $\mathbf{n}$ is known, the curvature tensor components are computed using~\eqref{eqn:kappa_t} whereas the geometrical curvature is derived from the sum of the three diagonal components of $\mathbf{K}$, i.e., $\kappa=-\left(K^{xx}+K^{yy}+K^{zz}\right)$.

\subsubsection{Calculation of normalization parameter $d_{th}$}\label{subsec:d_th}
Once $\mathcal{T}$ is known, the last step to obtain $H(\tilde{\mathbf{x}})$ is to compute the normalization parameter $d_{th}$. To this purpose, we impose the volume conservation at discrete level by setting that the volume integral of $\hat{H}$ over local grid cell is equal to the VoF field in that cell. Using the normalized Cartesian coordinate, this results in:
\begin{equation}
  \mathcal{I}: \int_0^1\int_0^1\int_0^1\hat{H}d\tilde{x}d\tilde{y}d\tilde{z}= \phi\mathrm{.}
  \label{eqn:int}
\end{equation}
As remarked in~\cite{ii2012interface}, an exact integration of~\eqref{eqn:int} is not possible while it exists for the one dimensional integration. Taking for example the $\tilde{x}$ direction for an exact integration, we get:
\begin{equation}
  \int_0^1\hat{H}(\tilde{x})d\tilde{x}=\dfrac{1}{2}\log\left[\tilde{x}+\dfrac{\log\left(\cosh\left(\beta_{th}\left(\mathcal{T}\left(\tilde{x},\tilde{y},\tilde{z}\right)\right)\right)\right)}{\beta_{th}\left(\partial\mathcal{T}_{\tilde{x}}/\partial\tilde{x}\right)}\right]_0^1\mathrm{.}
  \label{eqn:int_h}
\end{equation}
Note that in the integration of~\eqref{eqn:int_h}, we assume that the derivative of $\mathcal{T}$ with respect to $\tilde{x}$ just depends on $\tilde{y}$ and $\tilde{z}$. This is achieved by setting $\mathbf{c}=[0,1,1]$ and $\mathbf{c}^R=[1,1,0]$. In the other two directions, numerical integration should be performed and in this work we employ a two-point Gaussian quadrature method. Therefore, the integral $\mathcal{I}$ in equation~\eqref{eqn:int} results in:
\begin{equation}
  8\phi = \sum_{p=1}^4\left[\left(\tilde{x}+\dfrac{\log\left(\cosh\left(\beta_{th}\left(\mathcal{T}\left(\tilde{x},r_{p}(p),r_{m}(p)\right)+d_{th}\right)\right)\right)}{a_L^x}\right)\right]_0^1\mathbf{,}
  \label{eqn:int_x_ex}
\end{equation}
where $r_p=1+\sqrt{3}/2[-1,+1,-1,+1]$ and $r_m=1+\sqrt{3}/2[-1,-1,+1,+1]$. In general, the direction along which the exact integration is performed cannot be decided a-priori. On the other hand, a criterion for this choice is based on the magnitude of the normal vector components and, therefore, three cases are possible.

\subsubsection*{Case 1}
If, $|n^x| \geq (|n^y|,|n^z|)$, the exact integration is performed only along $x$ and we set $\mathbf{c}=[0,1,1]$ and $\mathbf{c}^R=[1,1,0]$ in equation~\eqref{eqn:qud_fun}. Therefore $\mathcal{I}$ becomes:
\begin{equation}
  8\phi = \sum_{p=1}^4\left[\left(\tilde{x}+\dfrac{\log\left(\cosh\left(\beta_{th}\left(\mathcal{T}\left(\tilde{x},r_{p}(p),r_{m}(p)\right)+d_{th}\right)\right)\right)}{a_L^x}\right)\right]_0^1\mathbf{.}
  \label{eqn:int_x}
\end{equation}

\subsubsection*{Case 2}
On the other hand, if $|n^y| \geq (|n^x|,|n^z|)$, the exact integration is performed only along $y$ and we set $\mathbf{c}=[1,0,1]$ and $\mathbf{c}^R=[0,1,1]$ in equation~\eqref{eqn:qud_fun}. Therefore, $\mathcal{I}$ becomes:
\begin{equation}
  8\phi = \sum_{p=1}^4\left[\left(\tilde{y}+\dfrac{\log\left(\cosh\left(\beta_{th}\left(\mathcal{T}\left(r_{p}(p),\tilde{y},r_{m}(p)\right)+d_{th}\right)\right)\right)}{a_L^y}\right)\right]_0^1\mathbf{.}
  \label{eqn:int_y}
\end{equation}

\subsubsection*{Case 3}
Finally, if $|n^z| \geq (|n^x|,|n^y|)$, the exact integration is performed only along $z$ and we set $\mathbf{c}=[0,0,1]$ and $\mathbf{c}^R=[1,0,1]$ in equation~\eqref{eqn:qud_fun}. Therefore, $\mathcal{I}$ results:
\begin{equation}
  8\phi = \sum_{p=1}^4\left[\left(\tilde{z}+\dfrac{\log\left(\cosh\left(\beta_{th}\left(\mathcal{T}\left(r_{p}(p),r_{m}(p),\tilde{z}\right)+d_{th}\right)\right)\right)}{a_L^z}\right)\right]_0^1\mathbf{.}
  \label{eqn:int_z}
\end{equation}
\\
\par
Depending on the three different cases, equations~\eqref{eqn:int_x} or~\eqref{eqn:int_y} and or ~\eqref{eqn:int_z} can be solved for the only unknown $d_{th}$. To this purposed, they can be re-written first as:
\begin{equation}
  4+\dfrac{1}{\beta_{th}a_L^{p}}\log\left(\dfrac{(AB^{-}D+1)(AB^{+}D+1)(AC^{-}D+1)(AC^{+}D+1)}{A^2(B^{-}D+1)(B^{+}D+1)(C^{-}D+1)(C^{+}D+1)}\right) = 8\phi\mathrm{,}
  \label{eqn:bef_quar}
\end{equation}
where $a_L^p$ is $a_L^x$, $a_L^y$ or $a_L^z$ according to the three cases above. Finally, equation~\eqref{eqn:bef_quar} can be further recast in a more convenient quartic equation:
\begin{align}
  0         &=\underbrace{A^2B^{-}B^{+}C^{-}C^{+}(A^2-Q)}_{\alpha_4}D^4\mathrm{,} \nonumber\\
     &+ \,\,  \underbrace{A^2(B^{-}B^{+}(C^{-}+C^{+})+(B^{-}+B^{+})C^{-}C^{+})(A-Q)}_{\alpha_3}D^3\mathrm{,} \nonumber\\
     &+ \,\,  \underbrace{A^2((B^{-}B^{+})(C^{-}+C^{+})+B^{-}B^{+}+C^{-}C^{+})(1-Q)}_{\alpha_2}D^2\mathrm{,} \nonumber\\
     &+ \,\,  \underbrace{A(B^{-}+B^{+}+C^{-}+C^{+})(1-AQ)}_{\alpha_1}D\mathrm{,} \nonumber\\
     &+ \,\,  \underbrace{(1-A^2Q)}_{\alpha_0}\mathrm{.}
  \label{eqn:quartic_1}
\end{align}
where $D=\exp(2\beta_{th}d_{th})$ while the constants $A$, $B^{\pm}$, $C^{\pm}$ and $Q$ are given by:
\begin{equation}
  \begin{cases}
    A &= \exp\left(2\beta_{th}a_L^p\right)\mathrm{,} \\
    B^{\pm} &= \exp(2\beta_{th}\mathbf{a}_{v}\cdot\mathbf{r}_{p,B})\mathrm{,} \\
    C^{\pm} &= \exp(2\beta_{th}\mathbf{a}_{v}\cdot\mathbf{r}_{p,C})\mathrm{,} \\
    Q &= \exp(4\beta_{th}a_L^p(2\phi-1))\mathrm{.} 
  \end{cases}
\end{equation}
The coefficients $a_L^p$ and the expressions for $\mathbf{a}_v$, $\mathbf{r}_{p,B}$ and $\mathbf{r}_{p,C}$ are reported in tables~\ref{tab:coeff_for_quar_1} and~\ref{tab:coeff_for_quar_2}, differentiated among the three cases previously described. Once the coefficients $\alpha_{i=1,4}$ of the quartic equation~\eqref{eqn:quartic_1} are computed, a solution for $D$ can be found. In the current work, we adopt the approach proposed in~\cite{ii2012interface}. Instead of computing all the four complex roots, we look for the real and positive solution, fulfilling the constraint given by equation~\ref{eqn:int}. 
\begin{table}[h!]
\centering
\begin{tabular}{l|cc}
\hline
\textit{Case} & $a_L^p$ & $\mathbf{a}_v$   \\
\hline
\textit{1} & $a_L^x$ & $[a_{Q,2}^z,a_{Q,2}^z,a_{Q,1}^{yz},a_L^y,a_L^z]^T$ \\
\textit{2} & $a_L^y$ & $[a_{Q,2}^x,a_{Q,2}^z,a_{Q,1}^{xz},a_L^x,a_L^z]^T$ \\
\textit{3} & $a_L^z$ & $[a_{Q,2}^x,a_{Q,2}^y,a_{Q,1}^{xy},a_L^x,a_L^y]^T$ \\
\hline
\end{tabular}
\caption{Coefficients $a_L^p$ and $\mathbf{a}_v$ (as given by equations~\eqref{eqn:q1_comp},~\eqref{eqn:q2_comp} and~\eqref{eqn:l_comp}) needed to compute $A$, $B^{\pm}$, $C^{\pm}$ and $D$ depending on the three cases.}
\label{tab:coeff_for_quar_1}
\end{table}
\begin{table}[h!]
\centering
\begin{tabular}{l|cc}
\hline
\textit{Case} & $\mathbf{r}_{p,B}$ & $\mathbf{r}_{p,C}$ \\
\hline
\textit{1} & $[(r_p^{\pm})^2,(r_p^{-})^2,(r_p^{\pm}r_p^{-}),r_p^{\pm},r_p^{-}]^T$ &                            
             $[(r_p^{-})^2,(r_p^{\pm})^2,(r_p^-r_p^{\pm}),r_p^{-},r_p^{\pm}]^T$ \\
\textit{2} & $[(r_p^{\pm})^2,(r_p^{-})^2,(r_p^{\pm}r_p^{-}),r_p^{\pm},r_p^{-}]^T$ & 
             $[(r_p^{-})^2,(r_p^{\pm})^2,(r_p^-r_p^{\pm}),r_p^{-},r_p^{\pm}]^T$ \\
\textit{3} & $[(r_p^{\pm})^2,(r_p^{-})^2,(r_p^{\pm}r_p^{-}),r_p^{\pm},r_p^{-}]^T$ & 
             $[(r_p^{-})^2,(r_p^{\pm})^2,(r_p^-r_p^{\pm}),r_p^{-},r_p^{\pm}]^T$ \\
\hline
\end{tabular}
\caption{Coefficients $\mathbf{r}_{p,B}$ and $\mathbf{r}_{p,C}$ needed to compute $B^{\pm}$ and $C^{\pm}$ depending on the three cases. Note that $r_p^+=1+\sqrt{3}/2$ and $r_p^-=1-\sqrt{3}/2$.}
\label{tab:coeff_for_quar_2}
\end{table}
To this purpose, we first write equation~\eqref{eqn:quartic_1} as:
\begin{equation}
  D^4+\gamma_3D^3+\gamma_2D+\gamma_1D+\gamma_0=0\mathrm{,}
  \label{eqn:quartic_2}
\end{equation}
with $\gamma_3=a_3/a_4$, $\gamma_2=a_2/a_4$, $\gamma_1=a_1/a_4$ and $\gamma_0=a_0/a_4$. Next, equation~\eqref{eqn:quartic_2} is recast in a quadratic form as:
\begin{equation}
  (D^2+\varepsilon_1x+\varepsilon_2)^2-(\varepsilon_3D+\varepsilon_4)^2=0\mathrm{,}
  \label{eqn:quartic_3}
\end{equation}
where $2\gamma_3=2\varepsilon_1$, $\gamma_2=\varepsilon_1^2+2\varepsilon_2-\varepsilon_3^2$, $\gamma_1=2\varepsilon_1\varepsilon_2-2\varepsilon_3\varepsilon_4$ and $\gamma_0=\varepsilon^2-\varepsilon_4^2=\gamma_0$. Finally, by introducing the variable $z=2\varepsilon_2$ and by comparing equation~\eqref{eqn:quartic_2} with equation~\eqref{eqn:quartic_3}, a cubic equation can be derived:
\begin{equation}
  z^3+\eta_2z^2+\eta_1z+\eta_0=0\mathrm{,}
  \label{eqn:cubic_1}
\end{equation}
with $\eta_2=-\gamma_2$, $\eta_1=\gamma_1\gamma_3-4\gamma_0$ and $\eta_0 = \gamma_0(4\eta_2-\eta_3^2)-\eta_1^2$. Equation~\eqref{eqn:cubic_1} can be easily solved with the Cardano's formula. Excluding the complex solutions, setting $\lambda=-\eta_2^2/9+\eta_1/3$, $\mu=2\eta_2^3/27-\eta_1\eta_2/3+\eta_0$ and $\Delta=\lambda^2+4\mu^3$, the real one $z_r$ is given by:
\begin{equation}
  z_r = \begin{cases}
      \left(\dfrac{-\lambda+\sqrt{\Delta}}{2}\right)^{1/3}-\left(\dfrac{+\lambda+\sqrt{\Delta}}{2}\right)^{1/3}-\dfrac{\eta_2}{3} \hspace{1 cm} \text{if $\Delta\geq 0$}\mathrm{,} \\
      2\sqrt{-\mu}\cos\left[\dfrac{1}{3}\tan^{-1}\left(\dfrac{-\Delta}{-\lambda}\right)\right]-\dfrac{\eta_2}{3} \hspace{2.43 cm} \text{if $\Delta< 0$}\mathrm{.} \\
  \end{cases}
\end{equation}
Once $z_r$ is known, the coefficients $\varepsilon_{i=1,4}$ of equation~\eqref{eqn:quartic_3} can be found: $\varepsilon_1=\gamma_1/2$, $\varepsilon_2=z_r/2$, $\varepsilon_4=\sqrt{\varepsilon_2^2-\gamma_0}$ and $\varepsilon_3=(-\gamma_3/2+\varepsilon_1\varepsilon_2)/\varepsilon_4$. \\
The last step requires the solution of equation~\eqref{eqn:quartic_3}. Once more, four solutions are possible, but since $D>0$ by numerical constrains, only one positive and real solution is acceptable. This can be easily computed as:
\begin{equation}
  D = \dfrac{-(\varepsilon_1-\varepsilon_2)+\sqrt{(\varepsilon_1-\varepsilon_2)^2-4(\varepsilon_2-\varepsilon_4)}}{2}\mathrm{,}
\end{equation}
from which, the normalization parameter $d_{th}$ can be evaluated simply as:
\begin{equation}
  d_{th} = \dfrac{1}{2\beta_{th}}\log(D)\mathrm{.}
\end{equation}

\subsection{Computation of the numerical flux}
The approximate expression of $H$ is also used for the calculation of the numerical fluxes in the interface advection step. These are evaluated as:
\begin{equation}
  f_{i\pm 1/2,j,k}^{x} = \dfrac{1}{\Delta y\Delta z}\int_{t^n}^{t^{n+1}}\left[\int_{\Delta y}\int_{\Delta z}\left.u H(\mathbf{x},t)\right|_{i\pm 1/2,j,k}^n\,\ dy\,\ dz\right] dt\mathrm{,}
  \label{eqn:fx}
\end{equation}
\begin{equation}
  f_{i,j\pm 1/2,k}^{y} = \dfrac{1}{\Delta x\Delta z}\int_{t^n}^{t^{n+1}}\left[\int_{\Delta x}\int_{\Delta z}\left.v H(\mathbf{x},t)\right|_{i,j\pm 1/2,k}^n\,\ dx\,\ dz\right] dt\mathrm{,}
  \label{eqn:fy}
\end{equation}
\begin{equation}
  f_{i,j,k\pm 1/2}^{z} = \dfrac{1}{\Delta x\Delta y}\int_{t^n}^{t^{n+1}}\left[\int_{\Delta x}\int_{\Delta y}\left.w H(\mathbf{x},t)\right|_{i,j,k\pm 1/2}^n\,\ dx\,\ dy\right] dt\mathrm{.}
  \label{eqn:fz}
\end{equation}
Note that the use of equations~\eqref{eqn:fx},~\eqref{eqn:fy} and~\eqref{eqn:fz} is impractical since they contain both temporal and spatial integrations. For a more simple evaluation, the time integral is 
replaced by a space integral performing a change of variable. Taking as an example the integral along $x$ and using the cell-centered coordinate system, we define
\begin{equation}
  \Delta\tilde{x}_{i\pm 1/2,j,k} = 
    \begin{cases}
      \left[1-\dfrac{\Delta t^{n+1}}{\Delta x}u_{i+1/2,j,k};1\right] &\text{$u_{i+1/2,j,k}\geq 0$}\mathrm{,} \\ 
      \left[0;-\dfrac{\Delta t^{n+1}}{\Delta x}u_{i+1/2,j,k}\right] &\text{$u_{i+1/2,j,k}<    0$}\mathrm{.} \\
    \end{cases}
\end{equation}
and we set $\Delta\tilde{y}=\Delta\tilde{z}=[0,1]$. Moreover, the indicator function in equation~\eqref{eqn:fx} is approximated with the VOF function at the current time step, $\phi^n$. Accordingly, $f_{i,j\pm 1/2,k}^{x}$ can be computed as:
\begin{equation}
  f_{i\pm 1/2,j,k}^{x} = 
    \begin{cases}
      \displaystyle{+\Delta\tilde{x}\int_{\Delta\tilde{x}_{i+1/2}}\int_{\Delta\tilde{y}}\int_{\Delta\tilde{z}}\hat{H}_{i,j,k}^{x,n}(\phi^n) d\tilde{V} \hspace{0.5 cm} \text{$u_{i+1/2,j,k}\geq 0$}}\mathrm{,} \\
      \displaystyle{-\Delta\tilde{x}\int_{\Delta\tilde{x}_{i-1/2}}\int_{\Delta\tilde{y}}\int_{\Delta\tilde{z}}\hat{H}_{i,j,k}^{x,n}(\phi^n) d\tilde{V} \hspace{0.5 cm} \text{$u_{i-1/2,j,k}< 0$}}\mathrm{.} \\
    \end{cases}
\end{equation}
where $d\tilde{V}=d\tilde{x}d\tilde{y}d\tilde{z}$. Likewise, $f_{i,j\pm 1/2,k}^{y,n}$ and $f_{i,j,k\pm 1/2}^{z,n}$ can be computed as:
\begin{equation}
  f_{i,j\pm 1/2,k}^{y} = 
    \begin{cases}
      \displaystyle{+\Delta\tilde{y}\int_{\Delta\tilde{y}_{i+1/2}}\int_{\Delta\tilde{y}}\int_{\Delta\tilde{z}}\hat{H}_{i,j,k}^{y,n}(\phi^x) d\tilde{V} \hspace{0.5 cm} \text{$v_{i,j+1/2,k}\geq 0$}}\mathrm{,} \\
      \displaystyle{-\Delta\tilde{y}\int_{\Delta\tilde{y}_{i-1/2}}\int_{\Delta\tilde{y}}\int_{\Delta\tilde{z}}\hat{H}_{i,j,k}^{y,n}(\phi^x) d\tilde{V} \hspace{0.5 cm} \text{$v_{i,j-1/2,k}< 0$}}\mathrm{.} \\
    \end{cases}
\end{equation}
\begin{equation}
  f_{i,j,k\pm 1/2}^{z} = 
    \begin{cases}
      \displaystyle{+\Delta\tilde{z}\int_{\Delta\tilde{z}_{i+1/2}}\int_{\Delta\tilde{x}}\int_{\Delta\tilde{z}}\hat{H}_{i,j,k}^{z,n}(\phi^y) d\tilde{V} \hspace{0.5 cm} \text{$w_{i,j,k+1/2}\geq 0$}}\mathrm{,} \\
      \displaystyle{-\Delta\tilde{z}\int_{\Delta\tilde{z}_{i-1/2}}\int_{\Delta\tilde{y}}\int_{\Delta\tilde{z}}\hat{H}_{i,j,k}^{z,n}(\phi^y) d\tilde{V} \hspace{0.5 cm} \text{$w_{i,j,k-1/2}< 0$}}\mathrm{.} \\
    \end{cases}
\end{equation}
with:
\begin{equation}
  \Delta\tilde{y}_{i,j\pm 1/2,k} = 
    \begin{cases}
      \left[1-\dfrac{\Delta t^{n+1}}{\Delta y}v_{i,j+1/2,k};1\right] &\text{$v_{i,j+1/2,k}\geq 0$}\mathrm{,} \\ 
      \left[0;-\dfrac{\Delta t^{n+1}}{\Delta y}v_{i,j+1/2,k}\right] &\text{$v_{i,j+1/2,k}<    0$}\mathrm{.} \\
    \end{cases}
\end{equation}
\begin{equation}
  \Delta\tilde{z}_{i,j,k\pm 1/2} = 
    \begin{cases}
      \left[1-\dfrac{\Delta t^{n+1}}{\Delta z}w_{i,j,k+1/2};1\right] &\text{$w_{i,j,k+1/2}\geq 0$}\mathrm{,} \\ 
      \left[0;-\dfrac{\Delta t^{n+1}}{\Delta z}w_{i,j,k+1/2}\right] &\text{$w_{i,j,k+1/2}<    0$}\mathrm{.} \\
    \end{cases}
\end{equation}

\subsubsection{Overall VoF algorithm}
Below, we report the overall VoF algorithm in the pseudocode~\ref{alg:vof_algo}. As a final remark, note that the entire algorithm has been described assuming that the first directional split is always oriented along $x$ (i.e., $x\rightarrow y\rightarrow z$. Nevertheless, this solution proves to be only first-order accurate in time. To improve the time accuracy of the solution, one possibility is to alternate the splitting direction as suggested in~\cite{strang1968construction}. 
\begin{algorithm}
  \caption{Overall VoF algorithm}\label{alg:vof_algo}
  \begin{algorithmic}[1]
    \State Set $(u^x,u^y,u^z)=(u,v,w)$;
    \For{$p=x,y,z$}
      \State Compute the numerical fluxes $f^p$;
      \State Compute $\phi^{p}$;
      \State Set the boundary conditions on $\phi^p$;
      \If{$p!=z$}
         \State Using $\phi^p$, update $\mathbf{n}$ and $\kappa$ with the procedure described in~\ref{subsec:normal_kappa} and $d_{th}$ with the procedure in~\ref{subsec:d_th};
      \EndIf
    \EndFor
    \State Compute $\phi^{n+1}$ using $\phi^{x}$, $\phi^{y}$ and $\phi^{z}$.
    \State Using $\phi^{n+1}$, update $\mathbf{n}$ and $\kappa$ with the procedure described in~\ref{subsec:normal_kappa} and $d_{th}$ with the procedure in~\ref{subsec:d_th};
  \end{algorithmic}
\end{algorithm}

\bibliographystyle{elsarticle-num}
\bibliography{mybibfile}







\end{document}